\documentclass{aa}

\usepackage{graphics}
\usepackage{epsfig}

\def\bib{\bibitem{}}

\newcommand{\xia}{\overline{\xi}}
\newcommand{\rhob}{\overline{\rho}}

\newcommand{\gam}{\gamma}
\newcommand{\inta}{\int_{-i\infty}^{+i\infty}}
\newcommand{\pl}{\partial}
\newcommand{\beq}{\begin{equation}}
\newcommand{\eeq}{\end{equation}}
\newcommand{\lag}{\langle}
\newcommand{\rag}{\rangle}
\newcommand{\De}{{\cal D}}
\newcommand{\om}{\omega}

\newcommand{\kappamin}{\kappa_{\rm min}}
\newcommand{\kperp}{{\bf k}_{\perp}}
\newcommand{\kpar}{k_{\parallel}}
\newcommand{\phimu}{\tilde{\varphi}_{\mu}}
\newcommand{\Phimu}{\Phi_{\mu}}
\newcommand{\psimu}{\psi_{\mu}}
\newcommand{\ximu}{\xi_{\mu}}
\newcommand{\Gam}{\Gamma}
\newcommand{\tvarphi}{\tilde{\varphi}}
\newcommand{\Om}{\Omega_{\rm m}}
\newcommand{\Ol}{\Omega_{\Lambda}}
\newcommand{\kapthe}{\kappa_{\theta}}
\newcommand{\tphikaptheh}{\tvarphi_{\hat{\kappa}_{\theta}}}
\newcommand{\phikaptheh}{\varphi_{\hat{\kappa}_{\theta}}}
\newcommand{\kappah}{\hat{\kappa}}
\newcommand{\kaptheh}{\hat{\kappa}_{\theta}}
\newcommand{\xikaptheh}{\xi_{\hat{\kappa}_{\theta}}}
\newcommand{\Ikap}{I_{\kappa}}
\newcommand{\Dk}{\De k \theta}

\newcommand{\xikapthe}{\xi_{\kappa_{\theta}}}
\newcommand{\yskapr}{y_{s,\kaptheh}}
\newcommand{\kapsr}{\kappa_{\theta , s}}

\newcommand{\kapso}{\kappa_{\theta , s, {\rm m}}}

\newcommand{\zmax}{z_{\rm max}}
\newcommand{\kapb}{\overline{\kappa}}
\newcommand{\chimax}{\chi_{\rm max}}
\newcommand{\wb}{\overline{w}}
\newcommand{\Fsb}{\overline{F}_s}
\newcommand{\kappaminb}{\kapb_{\rm min}}
\newcommand{\xikappab}{\xi_{\kapb_{\theta}}}
\begin{document}
%
%
%
%
\renewcommand{\textfraction}{.01}
\renewcommand{\topfraction}{0.99}
\renewcommand{\bottomfraction}{0.99}
\setlength{\textfloatsep}{2.5ex}
\thesaurus{Sect.02 (12.03.4; 12.07.1; 12.12.1)}
\title{Statistical properties of the convergence due to weak gravitational lensing by non-linear structures}   
\author{Patrick Valageas}
\institute{Service de Physique Th\'eorique, CEA Saclay, 91191 Gif-sur-Yvette, 
France}
\date{Received / Accepted }
\maketitle
\markboth{P. Valageas: Statistical properties of the convergence}{P. Valageas: Statistical properties of the convergence}

\begin{abstract}

Density fluctuations in the matter distribution lead to distortions of the images of distant galaxies through weak gravitational lensing effects. This provides an efficient probe of the cosmological parameters and of the density field. In this article, we investigate the statistical properties of the convergence due to weak gravitational lensing by non-linear structures (i.e. we consider small angular windows $\theta \la 1'$). Previous studies have shown how to relate the second and third order moments of the convergence to those of the density contrast while models based on the Press-Schechter prescription provide an estimate of the tail of $P(\kappa)$. Here we present a method to obtain {\it an estimate of the full {\it p.d.f.} of the convergence $\kappa$}. It is based on a realistic description of the density field which applies to overdense as well as underdense regions. We show that our predictions agree very well with the results of N-body simulations for the convergence. This could allow one to derive the cosmological parameters $(\Om,\Ol)$ as well as the full {\it p.d.f.} $P(\delta_R)$ of the density contrast itself in the non-linear regime from observations. Hence this gives a very powerful tool to constrain scenarios of structure formation.

\end{abstract}


\section{Introduction}

One important goal of cosmology is to obtain the properties of the dark matter density field which eventually led to the formation of galaxies, clusters and other astrophysical objects we observe today. The traditional method to obtain some observational constraints on the distribution of matter is to build large surveys of galaxies, which map the distribution of light, and to use a theoretical model to link these galaxies to the density field. However, this indirect technique presents the disadvantage to introduce significant uncertainties due to the relation between mass and light one needs to introduce. Hence it is clearly important to build other independent methods which probe in a more direct fashion the distribution of matter. Such a tool is provided by the gravitational distortion of light rays coming from very distant sources. Indeed, the density fluctuations along (or close to) the line of sight amplify and shear the images of distant galaxies. Then, one can measure for instance the shear induced by weak gravitational lensing from the observed ellipticities of these galaxies, if we assume that the source galaxy ellipticities are uncorrelated. Since such effects are purely gravitational they probe the total matter content of the universe, dark matter as well as baryonic matter. In particular, the statistical properties of the convergence or the shear induced by gravitational lensing can be directly linked to the characteristics of the density field. Many authors have already considered this problem, focusing especially on the second and third order moments in the quasi-linear regime relevant for large angular windows (e.g., Blandford et al.1991; Miralda-Escude 1991; Kaiser 1992; Bernardeau et al.1997; van Waerbeke et al.1999). The advantage of such large angular windows ($\theta \ga 10'$) is that one can use rigorous perturbative results to describe the properties of the density field (e.g., Bernardeau 1994). However, smaller angular scales also present a strong interest as they probe the non-linear regime which is directly linked to astrophysical objects like galaxies, Lyman-$\alpha$ clouds,... Thus, they could provide a check on the scenarios used to describe structure formation as well as the building of galaxies. Moreover, going to smaller scales is also an observational advantage since the signal is much easier to measure because it becomes larger while the systematics remain basically the same (e.g., van Waerbeke et al.1999).

Unfortunately, the non-linear regime is much more difficult to handle theoretically. Thus, previous theoretical studies were restricted to second or third order moments, using fits to the results of N-body simulations for the non-linear evolution of the power-spectrum (Jain \& Seljak 1997; Hui 1999), or described the density field as a collection of virialized halos, obtained from the Press-Schechter prescription (Press \& Schechter 1974), which allows one to model the tail of the distribution of the convergence (e.g., Porciani \& Madau 1999). However, these works did not provide a realistic description of the full probability distributions of the convergence. Indeed, this requires a consistent model for the density field which can describe underdense as well as overdense regions. On the other hand, several numerical studies have been performed using N-body simulations (e.g., Jain et al.1999). In this article, extending an earlier work relevant for point sources (Valageas 1999a), we present an analytic method to obtain the {\it p.d.f.} of the convergence. This is based on a {\it scaling model} developed in Balian \& Schaeffer (1989) (see also Bernardeau \& Schaeffer 1992; Valageas \& Schaeffer 1997) which has been seen to provide a good description of the non-linear density field through comparisons with numerical simulations (e.g., Valageas et al.2000; Bouchet et al.1991; Munshi et al.1999).

This article is organized as follows. In Sect.\ref{Properties of the density field} we briefly present the model we use for the density field and we introduce a few statistical tools. Next, in Sect.\ref{Distortions induced by weak gravitational lensing} we describe how we obtain the {\it p.d.f.} of the convergence $\kappa$, smoothed on small angular scales ($\theta \la 1'$). We consider the case where all sources are at the same redshift $z_s$ as well as a broad redshift distribution. Then, in Sect.\ref{Numerical results} we describe our numerical predictions, which we compare to available results from N-body simulations, for three cosmologies. Finally, in Sect.\ref{Dependence on the redshift of the source} we briefly consider the dependence of the weak gravitational lensing effects on the redshift of the sources.

\section{Properties of the density field}
\label{Properties of the density field}

In order to obtain the characteristics of the weak gravitational lensing effects we need a detailed description of the density field which gives rise to these distortions. Hence we briefly recall here the formalism we use to describe the density fluctuations. See Balian \& Schaeffer (1989) for the details of our approach (see also Valageas \& Schaeffer 1997; Valageas et al.2000). We shall use the same methods to obtain the properties of the weak gravitational lensing effects.

Our main tool to obtain the probability distributions of the various quantities we consider will be the generating functions defined from the moments or the cumulants of these variables, which are widely used in statistics (see also Balian \& Schaeffer 1989). Thus, the probability distribution $P(\mu)$ of a random variable $\mu$ can be derived from the generating function $\psimu(y)$ (similar to a Fourier transform) defined from the moments $\lag \mu^p\rag$ (provided they are finite) by: 
\beq
\psimu(y) = \sum_{p=0}^{\infty} \frac{(-1)^p}{p!} \; \lag\mu^p\rag \; y^p
\label{psimu}
\eeq
since we have:
\beq
P(\mu) = \inta \frac{dy}{2\pi i} \; e^{\mu y} \; \psimu(y)
\label{Pmupsimu}
\eeq
Moreover, it is often useful to introduce the generating function $\Phimu(y)$ defined from the cumulants $\lag\mu^p\rag_c$ by:
\beq
\Phimu(y) = \sum_{p=1}^{\infty} \frac{(-1)^p}{p!} \; \lag\mu^p\rag_c \; y^p
\label{Phimu}
\eeq
which verifies:
\beq
\Phimu(y) = \ln \left[ \psimu(y) \right]
\label{Phimupsimu}
\eeq
In the following, we consider variables such that their mean is zero. Then it is convenient to define a new generating function $\phimu(y)$ by:
\beq
\phimu(y) = \sum_{p=2}^{\infty} \frac{(-1)^{p-1}}{p!} \; \frac{\lag\mu^p\rag_c}{\lag\mu^2\rag^{p-1}} \; y^p \hspace{0.3cm} , \hspace{0.3cm} \lag\mu\rag=0
\label{phimu}
\eeq
Note that in this case $\lag\mu^2\rag=\lag\mu^2\rag_c$ since $\lag\mu\rag=0$. Then, using (\ref{Phimupsimu}) and (\ref{Pmupsimu}) we obtain:
\beq
\psimu(y) = e^{-\phimu(y \ximu) / \ximu} \hspace{0.4cm} \mbox{with} \hspace{0.4cm} \ximu = \lag\mu^2\rag
\eeq
and:
\beq
P(\mu) = \inta \frac{dy}{2\pi i \ximu} \; e^{[\mu y-\phimu(y)] / \ximu}
\label{Pmuphimu}
\eeq

This formalism allows us to determine for instance the probability distribution $P(\delta_R)$ of the density contrast $\delta_R$ within spherical cells of volume $V$, radius $R$:
\beq
\delta_R = \int_V \frac{d^3r}{V} \; \delta({\bf r}) \hspace{0.3cm} \mbox{with} \hspace{0.3cm} \delta = \frac{\rho-\rhob}{\rhob}
\eeq
in terms of the many-body correlation functions $\xi_p({\bf r}_1,...,{\bf r}_p)$ (here $\rhob$ is the mean density of the universe). Indeed, from the definition of the correlation functions (e.g., Peebles 1980) we can write the cumulants $\lag\delta_R^{\;p}\rag_c$ as:
\beq
p \geq 2 \; : \;\; \lag\delta_R^{\;p}\rag_c = \xia_p = \int_V \frac{d^3r_1 ... d^3r_p}{V^p} \; \xi_p ({\bf r}_1,...,{\bf r}_p)
\label{xip}
\eeq
Next, we define the parameters $S_p$ and the generating function $\tvarphi(y)$ which corresponds to (\ref{phimu}) by:
\beq
\tvarphi(y) = \sum_{p=2}^{\infty} \frac{(-1)^{p-1}}{p!} \; S_p \; y^p \hspace{0.3cm} , \hspace{0.3cm} S_p = \frac{\xia_p}{\xia_2^{\; p-1}} \hspace{0.2cm} , \hspace{0.2cm} S_2 = 1
\label{phiy}
\eeq
Then, from (\ref{Pmuphimu}) one obtains (White 1979; Balian \& Schaeffer 1989):
\beq
P(\delta_R) = \inta \frac{dy}{2\pi i \xia} \; e^{[\delta_R y - \tvarphi(y)] /\xia}
\label{Pdelta}
\eeq
where we note the mean correlation $\xia_2$ at scale $R$ as $\xia$. The interest of the generating function $\tvarphi(y)$ is that in the quasi-linear regime the parameters $S_p$ remain finite in the limit $\xia \rightarrow 0$ so that one can estimate $P(\delta_R)$ from (\ref{Pdelta}) using the function $\tvarphi(y)$ obtained by rigorous perturbative methods for $\xia \rightarrow 0$ (Bernardeau 1994). Moreover, in the non-linear regime which we consider in this article the stable-clustering ansatz (Peebles 1980) also implies that the parameters $S_p$ reach a finite limit in the highly non-linear regime $\xia \gg 1$ for an initial power-spectrum which is a power-law (e.g., Valageas \& Schaeffer 1997). In this case, once $\tvarphi(y)$ is measured at one scale and time in the highly non-linear regime, the density probability distribution $P(\delta_R)$ can be obtained for any time and scale in the non-linear regime provided that one knows the behaviour of $\xia$. Note that the stable-clustering ansatz provides the behaviour of the correlation functions in the highly non-linear regime since it means that at these scales one has (Peebles 1980):
\beq
\xi_p(\lambda {\bf r}_1,...,\lambda {\bf r}_p ;a) = a^{3(p-1)} \;
\lambda^{-\gam(p-1)} \; \hat{\xi}_p({\bf r}_1,...,{\bf r}_p)
\label{scal1}
\eeq
where $a(t)$ is the scale-factor and $\gam$ is the (local) slope of the two-point correlation function. Note that for an initial linear power-spectrum which is a power-law $P(k) \propto k^n$ we have if stable-clustering is valid:
\beq
\gam = \frac{3(3+n)}{5+n}
\label{gamma}
\eeq
The consequences of the scaling-laws (\ref{scal1}) were studied in details in Balian \& Schaeffer (1989). In practice, one does not directly measure $\tvarphi(y)$ itself from numerical simulations or observations but the {\it p.d.f.} $P(\delta_R)$ from which it can be derived using (\ref{Pdelta}). More precisely, in the highly non-linear regime one considers the variable $x$ and the generating function $\varphi(y)$ defined by:
\beq
x = \frac{1+\delta_R}{\xia} \hspace{0.3cm} , \hspace{0.3cm} \varphi(y) = y + \tvarphi(y)
\label{defx}
\eeq
Then, using (\ref{Pdelta}), for sufficiently ``large'' density contrasts the {\it p.d.f.} $P(\delta_R)$ can be written as (Balian \& Schaeffer 1989):
\beq
\xia \gg 1 \; , \; (1+\delta_R) \gg \xia^{\;-\om/(1-\om)} \; : \; P(\delta_R) = \frac{1}{\xia^{\;2}} \; h(x)
\label{Phx}
\eeq
where the scaling function $h(x)$ is the inverse Laplace transform of $\varphi(y)$:
\beq
h(x) =  -\inta \frac{dy}{2 \pi i} \; e^{xy} \; \varphi(y)
\label{hphi}
\eeq
which obeys:
\beq
p \geq 1 \; : \;\; S_p = \int_0^{\infty}  x^p h(x) \; dx \hspace{0.3cm} , \hspace{0.3cm}  S_1=S_2=1  
\label{Sphx}
\eeq
In (\ref{Phx}) the exponent $\om$ comes from the behaviour of $\varphi(y)$ at large $y$, see (\ref{phiom}). Note that for large $\xia$ the range where (\ref{Phx}) is valid extends from very low densities (and negative $\delta_R$) up to infinity. Thus, one can obtain $h(x)$ from the counts-in-cells statistics measured in simulations. Then, one can derive $\varphi(y)$ from $h(x)$ since (\ref{hphi}) can be inverted as:
\beq
\varphi(y) = \int_0^{\infty} \; \left( 1 - e^{-xy} \right) \; h(x) \; dx
\label{phih}
\eeq
This also gives $\tvarphi(y)$ from (\ref{defx}). We recall in App.A the behaviour of $P(\delta_R)$ obtained for simple models for $h(x)$ and $\varphi(y)$ which are consistent with numerical results. The function $h(x)$ has been measured in the non-linear regime for various power-spectra (CDM models and power-laws) by several authors (Valageas et al.2000; Bouchet et al.1991; Colombi et al.1997; Munshi et al.1999). In particular, although Colombi et al.(1996) found a small scale-dependence other authors found that the numerical results were consistent with $h(x)$ being scale-independent in the non-linear regime. Thus, in the following we shall use the scaling function $h(x)$ obtained by Valageas et al.(2000) for $n=-2$. Note that $h(x)$ depends on the power-spectrum and it must be obtained from numerical simulations since there is no known method to derive analytically $h(x)$ (see App.A). In our case, we choose the scaling function measured for $n=-2$ because as we shall see in Sect.\ref{Dependence on scale of the contribution to weak gravitational lensing} and in Fig.\ref{figDelta2k} most of the contributions to the weak lensing effects we study in this article come from scales where $n \simeq -2$. As described in Valageas (1999b) this model provides the simplest realistic description of the non-linear density field which takes into account substructures and underdense as well as overdense regions.

Then, to obtain the properties of the non-linear density field we only need to model the evolution of the two-point correlation function $\xia(R,z)$ (or the non-linear power-spectrum) from the initial linear power-spectrum. To this order we use the fits given by Peacock \& Dodds (1996) which give the non-linear power-spectrum $P(k)$ from its linear counterpart. Note that the evolution of $\xia$ measured in numerical simulations (e.g., Valageas et al.2000; Peacock \& Dodds 1996) is consistent with the stable-clustering ansatz.

\section{Distortions induced by weak gravitational lensing}
\label{Distortions induced by weak gravitational lensing}

\subsection{Shear tensor}
\label{Shear tensor}

The gravitational lensing effects produced by density fluctuations along the trajectory of a photon lead to an apparent displacement of the source and to a distortion of the image. Thus, light coming from a direction ${\vec \theta}$ is deflected by a small angle $\delta {\vec \theta}$. However, the observable quantities are not the displacements $\delta {\vec \theta}$ themselves but the distortions induced by these deflections, which are given by the symmetric shear matrix (e.g., Jain \& Seljak 1997):
\beq
\Phi_{i,j} \equiv \frac{\pl \delta \theta_{i}}{\pl \theta_{j}} = -2 \int_0^{\chi_s} d\chi \; \frac{\De(\chi) \De(\chi_s-\chi)}{\De(\chi_s)} \; \nabla_i \nabla_j \phi(\chi)
\label{Phi}
\eeq
Here $\chi$ is the radial comoving coordinate (and $\chi_s$ corresponds to the redshift $z_s$ of the source):
\beq
d\chi = \frac{\frac{c}{H_0} \; dz}{\sqrt{\Ol+(1-\Om-\Ol)(1+z)^2+\Om(1+z)^3}}
\label{chi}
\eeq
while the angular distance $\De$ is defined by:
\beq
\De(z) = \frac{c / H_0}{\sqrt{1-\Om-\Ol}} \sinh \left( \sqrt{1-\Om-\Ol} \; \frac{H_0 \chi}{c} \right)
\label{De}
\eeq
The gravitational potential $\phi$ is related to the density fluctuations $\delta$ by Poisson's equation:
\beq
\Delta \phi = \frac{3}{2} \Om \; \frac{H_0^2}{c^2} \; (1+z) \; \delta
\label{Poisson}
\eeq
In (\ref{Phi}) we used the weak lensing approximation: the derivatives $\nabla_i \nabla_j \phi(\chi)$ of the gravitational potential are computed along the unperturbed trajectory of the photon. This assumes that the components of the shear tensor are small but the density fluctuations $\delta$ can be large (Kaiser 1992). The shear tensor $\Phi_{i,j}$ is usually decomposed into its trace $\kappa$ and the shear components $\gam_1$, $\gam_2$, defined by:
\beq
\kappa = - \frac{\Phi_{1,1}+\Phi_{2,2}}{2} 
\label{defkappa}
\eeq
and
\beq
\gam_1 = - \frac{\Phi_{1,1}-\Phi_{2,2}}{2} \hspace{0.2cm} , \hspace{0.2cm} \gam_2 = - \Phi_{1,2} \hspace{0.2cm} , \hspace{0.2cm} \gam = \gam_1 + i \; \gam_2
\label{defgam}
\eeq
Moreover, the magnification $\mu$ of the source is given by:
\beq
\mu = \frac{1}{(1-\kappa)^2 - |\gam|^2}
\label{mukappa1}
\eeq
so that for small values of the shear components we have: 
\beq
\kappa \ll 1 \; : \; \mu = 1+2\kappa
\label{mukappa2}
\eeq
Thus the goal of this article is to obtain the probability distribution of the component $\kappa$ of the shear tensor, as defined in (\ref{Phi}). This applies to point sources like supernovae or weak gravitational lensing effects which are filtered on small angular scales ($\theta \la 1'$) since we only consider here the non-linear regime.

\subsection{Convergence $\kappa$}
\label{Convergence kappa}

The probability distribution $P(\mu)$ of the magnification $\mu$ as defined in (\ref{mukappa2}) without smoothing was already obtained in Valageas (1999a). This directly gives the probability distribution $P(\kappa)$ of the convergence $\kappa$. Here we extend this method to the case of a finite smoothing angle $\theta$. This allows us to check the validity of our treatment through comparisons with available results from N-body simulations.

Using (\ref{Phi}) and (\ref{Poisson}) one can show (Bernardeau et al.1997; Kaiser 1998) that the convergence along a given line of sight is:
\beq
\kappa \simeq \frac{3\Om}{2} \int_0^{\chi_s} d\chi \; w(\chi,\chi_s) \; \delta(\chi)
\label{kappa}
\eeq
with:
\beq
w(\chi,\chi_s) = \frac{H_0^2}{c^2} \; \frac{\De(\chi) \De(\chi_s-\chi)}{\De(\chi_s)} \; (1+z)
\label{w}
\eeq
where $z$ corresponds to the radial distance $\chi$. Thus the convergence $\kappa$ can be expressed in a very simple fashion as a function of the density field. First, we can see from (\ref{kappa}) that {\it there is a minimum value $\kappamin(z_s)$ } for the convergence of a source located at redshift $z_s$, which corresponds to an ``empty'' beam between the source and the observer ($\delta=-1$ everywhere along the line of sight):
\beq
\kappamin = - \frac{3\Om}{2} F_s(\chi_s) 
\label{kappamin}
\eeq
with
\beq
F_s(\chi_s) = \int_0^{\chi_s} d\chi \; w(\chi,\chi_s)
\label{Fs}
\eeq
Next, we define the ``normalized'' convergence $\kappah$ by:
\beq
\kappah = \frac{\kappa}{|\kappamin|}
\label{kappah}
\eeq
which obeys $\kappah \geq -1$. If one smoothes the observations with a top-hat window in real space of small angular radius $\theta$ one rather considers the mean ``normalized'' convergence $\kaptheh$:
\beq
\kaptheh = \int_0^{\theta} \frac{d^2 \zeta}{\pi \theta^2} \int_0^{\chi_s} d\chi \; \frac{w(\chi,\chi_s)}{F_s(\chi_s)} \; \delta \left[ \chi, \De(\chi) \; {\vec \zeta} \right]
\label{kapthe}
\eeq
Here ${\vec \zeta}$ is a vector in the plane perpendicular to the line of sight (we restrict ourselves to small angular windows) over which we integrate within the disk $|{\vec \zeta}| \leq \theta$ (we note this by the short notation $\int_0^{\theta}$). Thus $\chi$ is the radial coordinate while $\De \; {\vec \zeta}$ is the two-dimensional vector of transverse coordinates.

\subsection{Probability distribution $P(\kappa)$}
\label{Real-space method}

Next, in order to obtain the {\it p.d.f.} $P(\kaptheh)$ of the convergence we simply need to derive the cumulants $\lag \kaptheh^p \rag_c$. This will provide the parameters $S_{\kaptheh,p}$ and the generating function $\tphikaptheh(y)$ in a fashion similar to (\ref{phiy}) and (\ref{Pdelta}). Indeed, from (\ref{kapthe}) we have $\lag \kaptheh \rag=0$. As noticed by several authors (e.g., Bernardeau et al.1997; Hui 1999; Munshi \& Coles 1999) the cumulants $\lag \kaptheh^p \rag_c$ can be simply expressed in terms of the $p-$point correlation functions of the density field through (\ref{kapthe}). Thus, we write:
\beq
\begin{array}{l} {\displaystyle \lag\kaptheh^p\rag_c = \lag \int_0^{\theta} \prod_{i=1}^p \frac{d^2 \zeta_i}{\pi \theta^2} \; \int_0^{\chi_s} \prod_{i=1}^p d\chi_i }  \\ \\ {\displaystyle \hspace{2.5cm} \times \; \prod_{i=1}^p \; \frac{w(\chi_i,\chi_s)}{F_s} \; \delta \left( \chi_i, \De_i \; {\vec \zeta}_i \right) \; \rag_c } 
\end{array}
\label{kappap1}
\eeq
From the definition of the correlation functions (Peebles 1980):
\beq
\xi_p \left( {\bf x}_1,...,{\bf x}_p \right) = \lag \delta \left( {\bf x}_1 \right) ... \delta \left( {\bf x}_p \right) \rag_c
\label{xipdeltac}
\eeq
we obtain:
\[
\begin{array}{l} {\displaystyle \lag\kaptheh^p\rag_c = \int_0^{\chi_s} d\chi_1 \frac{w(\chi_1,\chi_s)}{F_s} \int_{-\chi_1}^{\chi_s-\chi_1} \prod_{i=2}^p d\chi_i \frac{w(\chi_1+\chi_i,\chi_s)}{F_s} }  \\ \\ {\displaystyle \hspace{1.8cm} \times \; \int_0^{\theta} \prod_{i=1}^p \frac{d^2 \zeta_i}{\pi \theta^2} \;\;\; \xi_p \left( \begin{array}{l} \chi_1 \\ \De_1 \; {\vec \zeta}_1 \end{array} , ... ,  \begin{array}{l} \chi_1+\chi_p \\ \De_p \; {\vec \zeta}_p \end{array} ;z \right) } 
\end{array}
\]
where we made the change of variables $\chi_i \rightarrow \chi_1 + \chi_i$ for $i \geq 2$. Since the correlation length (beyond which the many-body correlation functions are negligible) is much smaller than the Hubble scale $c/H(z)$ (where $H(z)$ is the Hubble constant at redshift $z$) only values of $\chi_i$ which obey $|\chi_i| \ll c/H(z)$ contribute to the integral over $\chi_i$ (for $i \geq 2$). Hence we have $w(\chi_1+\chi_i,\chi_s) \simeq w(\chi_1,\chi_s)$, $\De_i \simeq \De_1$ and we can push the integration boundaries over $\chi_i$ (for $i \geq 2$) to infinity. Thus, we get:
\beq
\begin{array}{l} {\displaystyle \lag\kaptheh^p\rag_c = \int_0^{\chi_s} d\chi \left( \frac{w(\chi,\chi_s)}{F_s} \right)^p \int_{-\infty}^{\infty} \prod_{i=2}^p d\chi_i } \\ \\  {\displaystyle \hspace{1.5cm} \times \int_0^{\theta} \prod_{i=1}^p \frac{d^2 \zeta_i}{\pi \theta^2} \;\;\; \xi_p \left( \begin{array}{l} 0 \\ \De \; {\vec \zeta}_1 \end{array} , ... ,  \begin{array}{l} \chi_p \\ \De \; {\vec \zeta}_p \end{array} ;z \right) }
\end{array}
\label{kappap2}
\eeq
where we used the fact that $\xi_p$ is invariant through the translation $(-\chi_1,0)$ over the $p$ points ${\bf x}_i$: $\chi_i \rightarrow \chi_i - \chi_1$. Although the points $( \chi_i, \De \; {\vec \zeta}_i )$ cover a cylinder rather than a sphere, we approximate the integral over the $p$-point correlation function by:
\beq
\int \prod_{i=2}^p d\chi_i \; \int_0^{\theta} \prod_{i=1}^p \frac{d^2 \zeta_i}{\pi \theta^2} \;\;\; \xi_p \;\; \simeq \;\; S_p \; \Ikap^{p-1}
\label{Ikappap}
\eeq
in a fashion similar to $\xia_p = S_p \; \xia^{\;p-1}$, see (\ref{xip}) and (\ref{phiy}), where we defined:
\beq
\Ikap = \int d\chi_2 \int \frac{d^2 \zeta_1}{\pi \theta^2} \frac{d^2 \zeta_2}{\pi \theta^2} \;\;\; \xi_2 \left( \begin{array}{l} 0 \\ \De \; {\vec \zeta}_1 \end{array} , \begin{array}{l} \chi_2 \\ \De \; {\vec \zeta}_2 \end{array} ; z \right)
\label{Ikap}
\eeq
We discuss in more details in App.B the approximation (\ref{Ikappap}) for the case of a tree-model for the $p-$point correlation functions. However, our approach only assumes that the stable-clustering ansatz is valid (i.e. the coefficients $S_p$ are constant with time at a given physical scale). In particular, the approximation (\ref{Ikappap}) should provide reasonable results even if the $p-$point correlation functions are not exactly given by a tree-model. Thus, using (\ref{Ikappap}) we write the cumulants (\ref{kappap2}) as:
\beq
\lag\kaptheh^p\rag_c = \int_0^{\chi_s} d\chi \; \left( \frac{w}{F_s} \right)^p \; S_p \; \Ikap^{p-1}
\label{kappap3}
\eeq
From the definition (\ref{phimu}) and (\ref{kappap3}) we obtain the generating function $\tphikaptheh(y)$:
\beq
\tphikaptheh(y) = \int_0^{\chi_s} d\chi \; \frac{\xikaptheh}{\Ikap} \; \tvarphi \left( y \frac{w}{F_s} \frac{\Ikap}{\xikaptheh} \right)
\label{tphikaptheh}
\eeq
where we introduced the variance:
\beq
\xikaptheh = \lag \kaptheh^2 \rag = \int_0^{\chi_s} d\chi \left( \frac{w}{F_s} \right)^2 \Ikap(z)
\label{xikaptheh}
\eeq
Throughout this article, $\tvarphi$ (as in the r.h.s. of (\ref{tphikaptheh})) and $\varphi$, without subscript, refer to the generating functions defined for the density field in (\ref{phiy}), (\ref{defx}) and App.A.

Note that the variance $\xikaptheh$ we obtain in (\ref{xikaptheh}) does not rely on the approximation (\ref{Ikappap}) and it is exact (within the weak lensing approximation). We only use (\ref{Ikappap}) to get an approximation for the higher-order parameters $S_{\kaptheh,p}$ with $p \geq 3$. From (\ref{Pmuphimu}) we obtain the {\it p.d.f.} $P(\kaptheh)$ and $P(\kapthe)$ as:
\beq
P(\kaptheh) = \inta \frac{dy}{2\pi i \xikaptheh} \; e^{[\kaptheh y-\tphikaptheh(y)] / \xikaptheh}
\label{Pkaptheh}
\eeq
and
\beq
P(\kapthe) = \frac{1}{|\kappamin|} \; P ( \kaptheh )
\label{Pkapthe}
\eeq
The relations (\ref{Pkaptheh}) and (\ref{tphikaptheh}) are the main results of this article. They allow us to {\it derive the properties of the convergence from the counts-in-cells statistics in a straightforward fashion}. The term $\Ikap$ defined in (\ref{Ikap}) can be written in a more convenient form as a function of the power-spectrum $P(k)$ defined by:
\beq
\lag \delta({\bf k}_1)  \delta({\bf k}_2) \rag =  P(k_1) \; \delta_D( {\bf k}_1 + {\bf k}_2 )
\eeq
Thus, using:
\beq
\xi_2( {\bf x} ) = \int d^3k \; e^{i {\bf k . x}} \; P(k) \hspace{0.4cm} , \hspace{0.4cm} {\bf k . x} = \kpar \chi + \kperp {\bf .} \De \; {\vec \zeta}
\eeq
where $\kpar$ is the component of ${\bf k}$ parallel to the line of sight while $\kperp$ is the two-dimensional vector formed by the components of ${\bf k}$ perpendicular to the line of sight, we obtain
\beq
\Ikap = \pi \int_0^{\infty} \frac{dk}{k} \; \frac{\Delta^2(k;z)}{k} \; W^2 \left( \Dk \right)
\label{IkapDelta}
\eeq
where we defined at redshift $z$:
\beq
\Delta^2(k;z) = 4 \pi k^3 P(k;z)
\label{Delta2k}
\eeq
and:
\beq
W \left( \Dk \right) = \frac{2}{\Dk} \; J_1 \left( \Dk \right)
\eeq
Here $J_1$ is the Bessel function of the first kind of order 1. Note that throughout this article $x$ and $k$ refer to comoving quantities. 

From (\ref{kappap3}) we can also obtain the coefficients $S_{\kapthe,p}$ defined in a fashion similar to (\ref{phiy}):
\beq
S_{\kapthe,p} = \frac{\lag \kapthe^p \rag_c}{\lag \kapthe^2 \rag^{p-1}} = | \kappamin |^{2-p} \; \frac{\lag \kaptheh^p \rag_c}{\lag \kaptheh^2 \rag^{p-1}}
\eeq
In particular, we get for the skewness $S_{\kapthe,3}$ the expression:
\beq
S_{\kapthe,3} = \frac{2}{3 \Om} \; S_3 \; \frac{\int d\chi \; w^3 \; \Ikap^2}{\left[ \int d\chi \; w^2 \; \Ikap \right]^2}
\label{S3kapthe}
\eeq
We can see from (\ref{S3kapthe}) that {\it $S_{\kapthe,3}$ is (almost) independent of the normalization of the power-spectrum, while it shows a strong dependence on the cosmological parameter $\Om$}. Thus it could be used to {\it measure $\Om$}, see Bernardeau et al.(1997) for such a study in the quasi-linear regime.

It is possible to get some important information about the generating function $\tphikaptheh(y)$ by direct inspection of (\ref{tphikaptheh}). First, we note that if we define the function $\phikaptheh(y)$ by:
\beq
\phikaptheh(y) = y + \tphikaptheh(y)
\label{phitphi}
\eeq
as in (\ref{defx}), then from (\ref{tphikaptheh}) we see that $\phikaptheh(y)$ and $\varphi(y)$ bear the same relation as $\tphikaptheh(y)$ and $\tvarphi(y)$:
\beq
\phikaptheh(y) = \int_0^{\chi_s} d\chi \; \frac{\xikaptheh}{\Ikap} \; \varphi \left( y \frac{w}{F_s} \frac{\Ikap}{\xikaptheh} \right)
\label{phikaptheh}
\eeq
Moreover, using (\ref{Pkaptheh}) the {\it p.d.f.} $P(\kaptheh)$ can be expressed as:
\beq
P(\kaptheh) = \inta \frac{dy}{2\pi i \xikaptheh} \; e^{[(1+\kaptheh) y-\phikaptheh(y)] / \xikaptheh}
\label{Pphikaptheh}
\eeq
In this article, we use (\ref{tphikaptheh}) and (\ref{Pkaptheh}), rather than (\ref{phikaptheh}) and (\ref{Pphikaptheh}), because for small values of the variance $\xikaptheh$ the {\it p.d.f.} $P(\kaptheh)$ still looks like a gaussian (except in the tails of the distribution). This is most clearly seen in (\ref{Pkaptheh}) since $\tphikaptheh(y) = -y^2/2 + ...$. Moreover, the use of $\tvarphi(y)$ ensures that there is no linear term in $y$ in the exponent in (\ref{Pkaptheh}). Indeed, a small numerical error in the linear term of $\phikaptheh(y)$ (which must satisfy $\phikaptheh(y) = y - y^2/2 + ...$) would show up as a non-zero mean $\lag \kaptheh \rag$ if one uses (\ref{Pphikaptheh}). However, for theoretical considerations (\ref{phikaptheh}) and (\ref{Pphikaptheh}) are more convenient. Thus, from (\ref{phikaptheh}) and (\ref{phiom}) we see that {\it $\phikaptheh(y)$ and $\varphi(y)$ have the same power-law behaviour for large $y$}:
\beq
\mbox{Re}(y) \rightarrow +\infty \; : \; \phikaptheh(y) \sim a_{\kaptheh} \; y^{1-\omega} \hspace{0.3cm} , \hspace{0.3cm} a_{\kaptheh} > 0
\label{phikapom} 
\eeq
In our case $\om \simeq 0.3$. Here we also used the fact that $\Ikap$ is always positive, as shown by (\ref{IkapDelta}). Then, from (\ref{phikapom}) and (\ref{Pphikaptheh}) we check that $P(\kaptheh)=0$ for $\kaptheh \leq -1$. Indeed, for $\kaptheh \leq -1$ we can push the integration path in (\ref{Pphikaptheh}) towards the right, Re$(y) \rightarrow +\infty$, so that the integral vanishes. Thus, {\it our approximation (\ref{Ikappap}) has preserved the fact that $P(\kapthe)=0$ for $\kapthe \leq \kappamin$}. Moreover, using (\ref{phikapom}) in (\ref{Pphikaptheh}) we obtain the behaviour of the cutoff of $P(\kapthe)$ for $\kapthe \rightarrow \kappamin$:
\beq
\begin{array}{l} {\displaystyle \kapthe \rightarrow \kappamin \; : } \\ \\  {\displaystyle P(\kapthe) \sim \exp \left[ - \; \frac{\om \;  a_{\kaptheh}^{1/\om}}{\xikaptheh} \left( \frac{\kapthe-\kappamin}{(1-\om)|\kappamin|} \right)^{-(1-\om)/\om} \right] } \end{array}
\label{Pcutom}
\eeq
where we did not write multiplicative power-law factors. Next, using (\ref{ys}) we see from (\ref{tphikaptheh}) or (\ref{phikaptheh}) that {\it $\tphikaptheh(y)$ and $\phikaptheh(y)$ have a singularity (branch cut) at $\yskapr$} given by:
\beq
\yskapr = \frac{y_s}{\displaystyle \max_{\chi} \left( \frac{w}{F_s} \frac{\Ikap}{\xikaptheh} \right)} \hspace{0.5cm} , \hspace{0.5cm} \yskapr < 0
\label{yskapr}
\eeq
The location of this singularity is important since in the numerical integration of (\ref{Pkaptheh}) one must make sure that the integration path does not cross the branch cut (i.e. the integration path crosses the real axis at a point $y$ such that $y > \yskapr$). Moreover, as can be seen from (\ref{Pkaptheh}) the existence of the singularity at $\yskapr$ implies that {\it $P(\kaptheh)$ shows an exponential tail for large $\kaptheh$}:
\beq
\kaptheh \gg 1 \; : \;\; P(\kaptheh) \sim e^{-|\yskapr| \; \kaptheh / \xikaptheh}
\eeq
so that $P(\kapthe)$ obeys:
\beq
\kapthe \gg |\kappamin| \; : \;\; P(\kapthe) \sim e^{- \kapthe / \kapsr}
\label{Pkapexp}
\eeq
with:
\beq
\kapsr = \frac{|\kappamin| \; \xikaptheh}{|\yskapr|} = \frac{3\Om}{2} \; x_s \; \max_{\chi} (w \Ikap)
\label{kapsr}
\eeq
Here we used $x_s = 1/|y_s|$ as in (\ref{has}). Note that the exponential cutoff of the {\it p.d.f.} $P(\kapthe)$ is stronger for low-density universes and a lower normalization of the power-spectrum $P(k)$ (through $\Ikap$) since $\kapsr$ gets smaller. Of course, this is due to the fact that in such cases there are fewer high-density massive objects (which are the ones which can produce a high convergence $\kapthe$). Note that (\ref{kapsr}) shows that the cutoff of the {\it p.d.f.} $P(\kapthe)$  strongly depends on the cutoff $x_s$ of the density contrast distribution, so that a measure of the tail of $P(\kapthe)$ provides a direct estimate of the tail of $P(\delta_R)$ (but see the discussion in Sect.\ref{Dependence on the redshift of the source}).

The case without smoothing is recovered in all previous expressions by taking the limit $\theta \rightarrow 0$. In particular, (\ref{tphikaptheh}) and (\ref{Pkaptheh}) are unchanged while in (\ref{IkapDelta}) the factor $W^2$ is equal to $1$. Of course, we could also perform the same calculation directly without smoothing, so that $\kappa$ only involves the integration along one line of sight, as in (\ref{kappa}), and we would recover these results.

\subsection{A convenient approximation}
\label{A convenient approximation}

We can check from (\ref{tphikaptheh}) that the expansion about $y=0$ of the generating function $\tphikaptheh(y)$ satisfies: $\tphikaptheh(y) = -y^2/2 + ...$, as implied by the definition (\ref{phimu}). Moreover, from (\ref{Fs}), (\ref{xikaptheh}) and (\ref{tphikaptheh}) we see that $\tphikaptheh(y)$ should be close to $\tvarphi(y)$ since the factor $(w \Ikap) / (F_s \xikaptheh)$ has typical values of order unity. Thus, a simple approximation is to use:
\beq
\tphikaptheh(y) \simeq \tvarphi(y)
\label{app1}
\eeq
which simplifies somewhat the numerical calculations. Note that the approximation (\ref{app1}) actually means that we use $S_{\kaptheh,p} \simeq S_p$, i.e.:
\beq
S_{\kapthe,p} \simeq |\kappamin|^{2-p} \; S_p
\label{Spapp}
\eeq
The approximation (\ref{app1}) also implies that $P(\kapthe)$ shows an exponential cutoff for large positive $\kaptheh$ with a parameter $\kapso$ similar to $\kapsr$ given by:
\beq
\kapso = \frac{3\Om}{2} \; x_s \; ( F_s \xikaptheh )
\label{kapso}
\eeq
In fact, using (\ref{Pdelta}), the approximation (\ref{app1}) means that the {\it p.d.f.} $P(\kaptheh)$ is directly given by the {\it p.d.f.} $P(\delta_R)$ of the density contrast as:
\beq
P(\kapthe) = \frac{1}{|\kappamin|} \; P \left( \delta_R \rightarrow \frac{\kapthe}{|\kappamin|} \; ; \; \xia \rightarrow \xikaptheh \right)
\label{PkapPdel}
\eeq
This clearly shows that the {\it p.d.f.} $P(\kaptheh)$ is an efficient tool to obtain the properties of the density field. Indeed, we shall check in Sect.\ref{pdf kappa} that the approximation (\ref{app1}) gives very good results and the relation (\ref{PkapPdel}) provides a straightforward method to derive $P(\delta_R)$ from $P(\kaptheh)$. Note however that even for $\xikaptheh < 1$ the {\it p.d.f.} $P(\delta_R)$ used in (\ref{PkapPdel}) corresponds to the non-linear regime (for $\theta \la 1'$). Hence it is not the {\it p.d.f.} of the density contrast measured at the time when $\xia = \xikaptheh$ since the values of the coefficients $S_p$ in the linear and non-linear regimes are different. 

We shall see in Sect.\ref{pdf kappa} that the approximation (\ref{app1}) gives very good results. In particular, we noticed above in (\ref{phikapom}) that $\tphikaptheh(y)$ and $\tvarphi(y)$ obey the same power-law behaviour for large $y$. However, the location of their singularity along the real axis is slightly different, see (\ref{yskapr}), and the exponent $\om_s$ of the power-law prefactor at large $x$ of their associated inverse Laplace transform $h(x)$, defined as in (\ref{hphi}), differs by a factor $1/2$, as discussed in Valageas (1999a) (see the system (43)). More precisely, we obtain from (\ref{yskapr}):
\beq
\frac{|y_s|}{ | \yskapr | } = \frac{ \int d\chi \; w \;\;\; \max( w \Ikap ) } {\int d\chi \; w^2 \; \Ikap} > 1
\label{ysiysm}
\eeq 
which means that the singularity $\yskapr$ is closer to 0 than $y_s$. This implies that the exponential cutoff of $P(\kapthe)$ is slightly smoother for the more accurate expression (\ref{tphikaptheh}) than for the approximation (\ref{app1}), see (\ref{kapsr}). Next, we noticed above that (\ref{app1}) implies that the skewness $S_{\kaptheh,3}$ of the normalized convergence $\kaptheh$ is given by $S_{\kaptheh,3} = S_3$ while the expression (\ref{S3kapthe}) gives:
\beq
S_{\kaptheh,3} = S_3 \;\; \frac{\int d\chi \; w \;\;\; \int d\chi \; w^3 \; \Ikap^2}{\left[ \int d\chi \; w^2 \; \Ikap \right]^2} > S_3
\label{S3S3mean}
\eeq
for the more accurate calculation (\ref{tphikaptheh}). The inequality is obtained from the Cauchy-Schwarz inequality applied to the scalar product $\lag f | g \rag$ defined by: $\lag f | g \rag = \int d\chi \; f . g$ for the functions $f=w^{1/2}$ and $g=w^{3/2} \Ikap$. Hence the skewness of the convergence given by the approximation (\ref{app1}) is slightly too small. The inequalities (\ref{ysiysm}) and (\ref{S3S3mean}) translate the fact that the approximation (\ref{app1}) corresponds to an average along the line of sight of the kernel $(w \Ikap)/(F_s \xikaptheh)$ which appears in (\ref{tphikaptheh}), hence to an average over the various {\it p.d.f.} of the weak lensing effect arising from different planes along the line of sight. On the other hand, the expression (\ref{tphikaptheh}) takes into account the variations of this kernel so that the {\it p.d.f.} of the total convergence is given by the exact convolution of the various {\it p.d.f.} associated to the elementary contributions due to the successive mass planes along the line of sight (which are discretized in numerical ray-tracing simulations). Then, the tails of the {\it p.d.f.} $P(\kapthe)$ are sensitive to the highest tails among these successive {\it p.d.f.}, as shown by the maximum which appears in (\ref{yskapr}). This implies that the tails of $P(\kapthe)$ given by (\ref{tphikaptheh}) are smoother than the cutoffs implied by the approximation (\ref{app1}). This leads to a smaller cutoff $\kapsr$ and a smaller skewness for this approximation, as shown in (\ref{ysiysm}) and (\ref{S3S3mean}).

\subsection{Redshift distribution of the sources}
\label{Redshift distribution of the sources}

In the previous sections we considered the case where all the sources are located at the same redshift $z_s$. This is convenient to compare our predictions with the results of numerical simulations and to study the dependence on redshift of the weak gravitational lensing effects. However, in practice one observes galaxies over a finite range of redshifts in order to get a sufficiently large number of sources to perform a statistical analysis. Hence we show in this section how our results can be applied to a ``broad'' redshift distribution of sources $n(z_s) dz_s$. Here we normalized $n(z_s)$ to unity:
\beq
\int_0^{\zmax} dz_s \; n(z_s) = 1
\label{nzs}
\eeq
where $\zmax$ is the maximum redshift of the sources. Then the mean convergence $\kapb$ observed in one direction on the sky is the average of the convergence $\kappa$ defined in (\ref{kappa}) over the redshift distribution of the sources:
\beq
\kapb = \int_0^{\zmax} dz_s \; n(z_s) \; \kappa(z_s)
\eeq
which gives:
\beq
\kapb = \frac{3\Om}{2} \int_0^{\zmax} dz_s \; n(z_s) \int_0^{\chi_s(z_s)} d\chi \; w(\chi,\chi_s) \; \delta(\chi)
\label{kapb1}
\eeq
The expression (\ref{kapb1}) can also be written:
\beq
\kapb = \frac{3\Om}{2} \int_0^{\chimax} d\chi \; \wb(\chi) \; \delta(\chi)
\label{kapb2}
\eeq
with:
\beq
\wb(\chi) = \int_{z(\chi)}^{\zmax} dz_s \; n(z_s) \; w\left( \chi,\chi_s(z_s) \right)
\label{wb}
\eeq
Thus, we obtain exactly the same expression as in (\ref{kappa}) with $\chi_s$ replaced by $\chimax$ and $w(\chi,\chi_s)$ by $\wb(\chi)$. Hence all our previous results remain valid, after we perform these two changes and $F_s$ is replaced by $\Fsb$ (compare with (\ref{Fs})) obtained from:
\beq
\Fsb = \int_0^{\chimax} d\chi \; \wb(\chi)
\label{Fsb}
\eeq
In particular, the relations (\ref{Pkaptheh}) and (\ref{tphikaptheh}) are still correct after we make these straightforward changes. Hence our results also apply to the {\it p.d.f.} $P(\kapb_{\theta})$ of the convergence observed from a broad redshift distribution of sources. Moreover, we noticed in Sect.\ref{A convenient approximation} that a convenient approximation to the generating function $\tphikaptheh(y)$ (defined for a redshift $z_s$ of the sources) is simply $\tvarphi(y)$, see (\ref{app1}). This is checked below in Fig.\ref{figPlkappa}. In the case of a broad redshift distribution of the sources we can still perform the same approximation (\ref{app1}) since we noticed that the relation (\ref{tphikaptheh}) still holds (with the changes of notations described above). In fact, this could be expected since $\tvarphi(y)$ is independent of $z_s$. This means that, to a good approximation, the full {\it p.d.f.} $P(\kapb_{\theta})$ of the convergence exhibits a specific {\it scaling property}. Indeed, it is described by a unique generating function $\tvarphi(y)$ (which characterizes the properties of the underlying density field) which is independent of the cosmological parameters, of the angular scale $\theta$ and of the redshift distribution of the sources, and by two numbers $\kappaminb$ and $\xikappab$ (which describe the lower bound and the variance of this {\it p.d.f.}) which contain all the dependence on the cosmology, on $\theta$ and on the redshift distribution of the sources. This holds as long as one probes the non-linear regime up to $\zmax$ (i.e. $\theta$ must be small enough). In particular, the two {\it p.d.f.} $P(\kaptheh)$ of the normalized convergence $\kaptheh$ obtained for two different redshifts $z_{s1}$, $z_{s2}$, of the sources and two angular windows $\theta_1$, $\theta_2$, chosen such that the normalized variances are the same, $\xi_{\kappah_{\theta 1}} = \xi_{\kappah_{\theta 2}}$, should nearly superpose.

Note that in the treatment presented in this section the redshift distribution of the sources is described by a uniform background population $n(z_s)$. In fact the quantity $n(z_s)$ will fluctuate from one line of sight to another and these variations $\delta n$ should be correlated with the density fluctuations along the line of sight, because the galaxies are expected to be correlated with the dark matter density field. As a consequence, in the expression (\ref{kapb2}) the product $n(z_s) . \delta(\chi)$ contains a term $\delta n(z_s) . \delta(\chi)$ which should be taken into account. However, since the kernel $w(\chi,\chi_s)$ vanishes for $\chi=\chi_s$ most of the weak lensing effects which distort the image of a source arise from density fluctuations located at cosmological distances ($\sim c/H(z)$) along the line of sight from this source. This supresses somewhat the contribution of the coupling between the fluctuations of the sources $\delta n(z_s)$ and of the density field $\delta(\chi)$. Then, the contributions due to the fluctuations of the sources only, which lead to terms like $\lag \delta n^2 \rag$, become relatively small if the number of sources is sufficiently large. However, these effects would certainly deserve a more detailed study but this is beyond the scope of this article.

\section{Numerical results}
\label{Numerical results}

We now describe the numerical results we get from the analytical tools detailed in the previous sections. In order to compare our predictions with the available results obtained from ray-tracing through N-body simulations we consider the three cosmological models defined in Tab.\ref{table1}: a standard CDM scenario (SCDM), a low-density open universe (OCDM) and a low-density flat universe with a non-zero cosmological constant ($\Lambda$CDM). Here $\Gam$ is the usual shape parameter of the power-spectrum. We use the fit given by Bardeen et al.(1986) for $P(k)$. We consider the weak lensing distortions which affect a population of sources at redshift $z_s=1$.

\begin{table}
\begin{center}
\caption{Cosmological models}
\label{table1}
\begin{tabular}{|r|lll|}\hline

  & SCDM & OCDM &  $\Lambda$CDM \\ \hline

$\Om$ & 1 & 0.3 & 0.3 \\
$\Ol$ & 0 &  0 &   0.7  \\

$H_0$ [km/s/Mpc] & 50 & 70 & 70 \\

$\sigma_8$ & 0.6 & 0.85 & 0.9 \\

$\Gam$ & 0.5 & 0.21 & 0.21 \\ \hline

\end{tabular}
\end{center}
\end{table}

\subsection{Magnitude of the convergence}
\label{Magnitude of the convergence}

\begin{figure}

\begin{picture}(230,440)
{\epsfxsize=23 cm \epsfysize=17 cm \put(-65,-25){\epsfbox{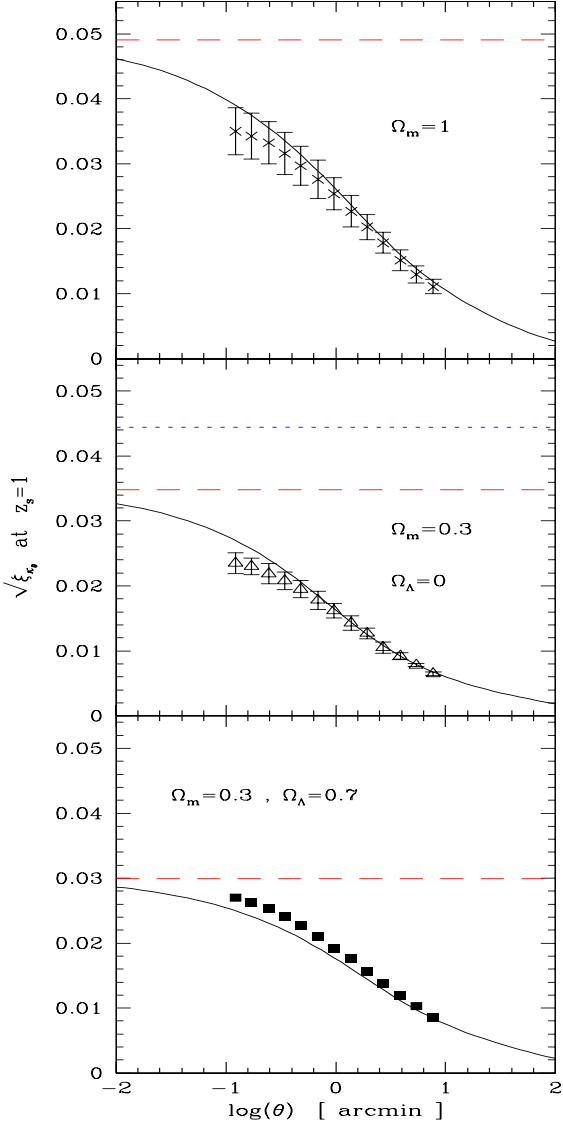}} }
\end{picture}

\caption{The magnitude of the convergence $\kapthe$ for a source at redshift $z_s=1$ for various angular windows $\theta$ and three cosmologies. The solid lines show the {\it rms} convergence $\sqrt{\xikapthe}$, the horizontal dashed lines show the {\it rms} convergence $\sqrt{\xi_{\kappa_0}}$ obtained without smoothing ($\theta=0$) and the horizontal dotted line in the middle panel shows $|\kappamin|$. For both flat universes $|\kappamin|$ is larger than $0.055$ ($0.12$ for the critical density universe and $0.064$ for the model with cosmological constant). The data points are the results of numerical simulations from Munshi \& Jain (2000).}
\label{figXikappa}

\end{figure}

First, we consider the {\it rms} convergence $\sqrt{\xikapthe} = |\kappamin| \sqrt{\xikaptheh}$ we obtain for various angular windows. Our results are displayed in Fig.\ref{figXikappa}.

We see that our results agree rather well with the N-body simulations (the points correspond to the average over several realizations while the error bars give the overall scatter, see Munshi \& Jain 2000). This is expected since our prediction for the variance $\xikapthe$ only relies on the weak lensing approximation (Kaiser 1992) and on the prescription given by Peacock \& Dodds (1996) to obtain the non-linear evolution of the power-spectrum. Since the latter is a fit to other N-body simulations, the agreement only shows that both sets of simulations are consistent. The deviations at small angular scales are due to the finite resolution, see Jain et al.(1999) for details. Although there remains a small discrepancy for the $\Lambda$CDM scenario it is certainly of the same order as the accuracy of the simulations. The main feature of Fig.\ref{figXikappa} is that {\it the variance is smaller for low-density universes}, despite their larger normalization $\sigma_8$ of the power-spectrum. Indeed, from (\ref{kappamin}) and (\ref{xikaptheh}) we can write the variance as:
\beq
\xikapthe = \left( \frac{3 \Om}{2} \right)^2 \; \int d\chi \; w^2  \pi \int \frac{dk}{k} \; \frac{\Delta^2(k)}{k} \; W^2 \left( \Dk \right)
\label{xikapthe}
\eeq
and the figure shows that the strong dependence on $\Om$ is more important than the variation with $\sigma_8$ (for the cosmologies we consider here, which have reasonable parameters). Of course, the variance is smaller for low-density universes (at fixed $\Delta(k)$) because in such cases there is less matter to produce weak lensing distortions. Since the difference between an ``empty'' beam ($\delta=-1$ everywhere along the line of sight) and the mean ($\delta=0$) is also smaller - it is proportional to $\Om$ as seen in (\ref{kappamin}) - the upper bound $|\kappamin|$ of the magnitude of negative deviations is lower than for the critical density universe. We can see from Fig.\ref{figXikappa} that an angular window $\theta=1'$ still removes some small scale power as compared to the case without smoothing (horizontal dashed lines). In particular, the latter, which corresponds to the observation of point sources like SNeIa (see Valageas 1999a), is still somewhat beyond the resolution of N-body simulations.

\subsection{Dependence on scale of the contribution to weak gravitational lensing}
\label{Dependence on scale of the contribution to weak gravitational lensing}

Next, in order to distinguish the contributions to weak gravitational lensing effects arising from different comoving scales we define the quantity:
\beq
\Delta^2_{\kapthe}(k) =  \frac{ {\displaystyle \int_0^{\chi_s} d\chi \; w^2 \; \pi \; \frac{\Delta^2(k)}{k} \; W^2 \left( \Dk \right) } }{ {\displaystyle \int_0^{\chi_s} d\chi \; w^2 \; \pi \int_0^{\infty} \frac{dk}{k} \; \frac{\Delta^2(k)}{k} } }
\label{Delta2ktheta}
\eeq
which measures the contribution from the comoving wavenumber $k$ to the variance of the convergence $\xi_{\kappa_{\theta}}$, see (\ref{xikaptheh}) and (\ref{IkapDelta}). Thus, $\Delta^2_{\kapthe}(k)$ integrated over $d\ln k$ is normalized to unity for the case without smoothing ($\theta=0$). Hence $\Delta^2_{\kapthe}(k)$ also shows the influence of the finite window $\theta$.  

\begin{figure}

\begin{picture}(230,440)
{\epsfxsize=23 cm \epsfysize=17 cm \put(-65,-25){\epsfbox{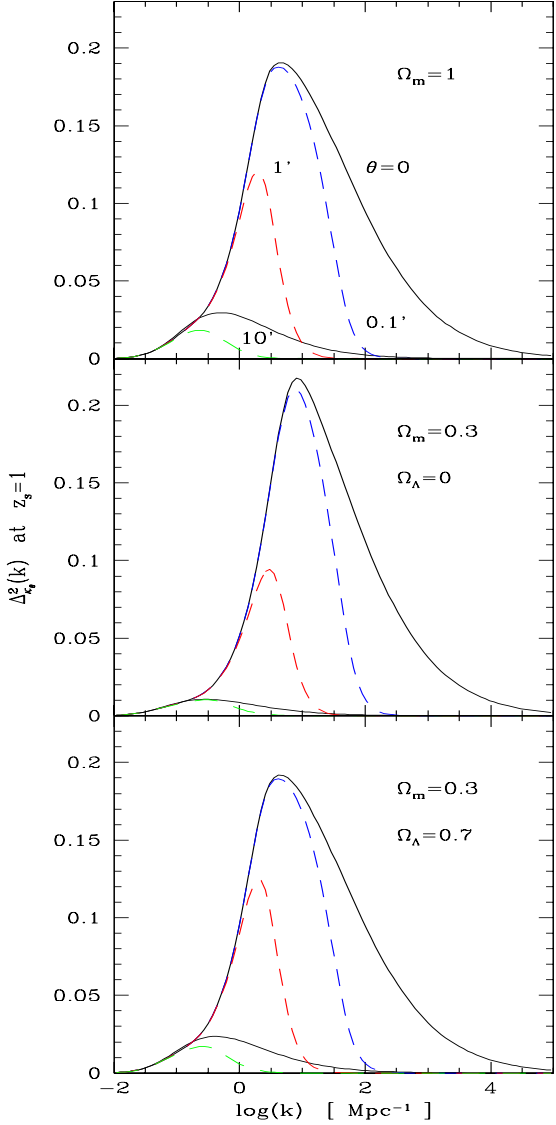}} }
\end{picture}

\caption{The contribution of various comoving wavenumbers $k$ to the convergence $\kapthe$ for a source at redshift $z_s=1$ for three cosmologies. The upper solid lines show $\Delta^2_{\kapthe}(k)$ for the case without smoothing ($\theta=0$). The dashed lines show the cases $\theta= 0.1'$, $\theta= 1'$ and $\theta=10'$. A larger angular window $\theta$ corresponds to a curve with a smaller upper cutoff in $k$ and a lower maximum. The lower solid curves show $\Delta^2_{\kappa_0}(k)$ (i.e. no smoothing) when the power-spectrum $P(k)$ is always taken to be the linear extrapolation.}
\label{figDelta2k}

\end{figure}

We present in Fig.\ref{figDelta2k} our results for three different angular windows, $\theta= 0.1'$, $\theta= 1'$ and $\theta=10'$ (shown by the dashed curves), as well as the case without smoothing (upper solid curve). As we can clearly see in the figure, {\it the effect of the finite window is to cut the contribution from small scales} (smaller than the size of the window), i.e. from high comoving wavenumbers $k$. Thus, even the smallest window $\theta= 0.1'$ cuts a significant part of the small-scale power which contributes to the case without smoothing. The latter is relevant for point sources, e.g. supernovae (see Valageas 1999a for a detailed study and the implications for the measure of the cosmological parameters). Since present numerical simulations already miss some power for $\theta=0.1'$ (Jain et al.1999) they would significantly underestimate the effect associated with point sources (in particular note the extended tail at high $k$). The lower solid lines in the figure correspond to the case without smoothing where we use the linear power-spectrum $P_L(k)$ instead of the actual power-spectrum $P(k)$. Thus, we see from the figure that for small angular windows $\theta \la 1'$ the non-linear evolution of the density fluctuations plays an important role and must be taken into account, as was already noticed by Jain \& Seljak (1997). This shows that the formalism we developed in Sect.\ref{Distortions induced by weak gravitational lensing} applies to $\theta \la 1'$ where the relevant scales are within the non-linear regime. For very large scales $\theta \ga 10'$ we could also use the same formalism but the parameters $S_p$ would be different (see Colombi et al.1997). The intermediate regime is rather difficult to model accurately since it corresponds to a transitory regime where the coefficients $S_p$ evolve (but see Hui 1999).

For $\theta= 0$ the characteristic scale which dominates the weak lensing distortions (shown by the maximum of the upper solid curve in the figure) is set by the shape of the initial power-spectrum itself. Thus, from (\ref{Delta2ktheta}) we see at once that this corresponds to the wavenumber $k_{\rm m}$ where the local slope of $\Delta^2(k)$ is $1$, i.e. $\Delta^2(k) \propto k$. From the definition (\ref{Delta2k}) this means $P(k) \propto k^{-2}$, i.e. $n=-2$ or $\xi(r) \propto r^{-1}$. In fact, we can see from (\ref{gamma}) that an initial power-spectrum $P_L(k) \propto k^{-2}$ keeps the same slope $n=-2$ in the highly non-linear regime. However, we see from the figure that $\Delta^2_{\kappa_0}(k)$ peaks at a value $k_{\rm m}$ somewhat larger than the wavenumber $k_{-2}$ where the local slope of the initial linear power-spectrum $P_L(k)$ is equal to $-2$, which corresponds to the peak of the lower solid curve. This is due to the form of the detailed evolution from the linear regime to the highly non-linear regime. However, we can check in the figure that for $\theta \la 10'$ {\it the main contribution to the weak lensing distortions comes from wavenumbers which are close to $k_{-2}$}. This justifies the use of (\ref{fithx}) which gives the parameters $S_p$ for the case $P(k) \propto k^{-2}$. In fact, the curvature of the actual power-spectrum might lead to a small modification of these coefficients $S_p$ but our approximation has the advantage of the simplicity and should provide a reasonable prescription. In order to improve our treatment, one would need to measure the function $\varphi(y)$ from the statistics of the counts-in-cells, as explained in Sect.\ref{Properties of the density field}, on the relevant scale for each cosmological scenario.

\subsection{Skewness}
\label{Skewness}

A convenient measure of the non-gaussianity of a peculiar {\it p.d.f.} is the skewness $S_3$. In particular, we can consider the skewness $S_{\kapthe,3}$ of the convergence which we obtained in Sect.\ref{Convergence kappa}. 

\begin{figure}

{\epsfxsize=8 cm \epsfysize=5.4 cm \epsfbox{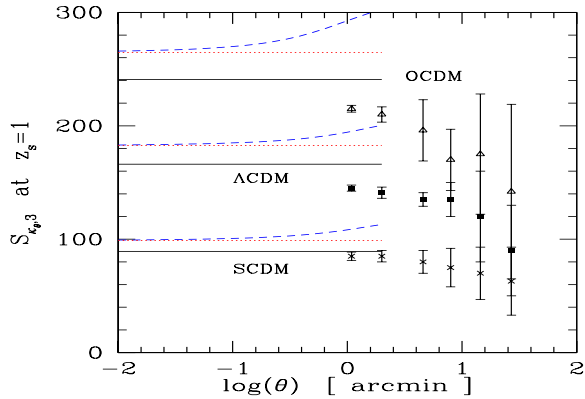} }

\caption{The skewness $S_{\kapthe,3}$ of the convergence for a source at $z_s=1$. As in Fig.1 we consider three cosmologies: open universe, low-density flat model and critical density universe, from top downto to bottom. The dashed curve corresponds to (\ref{S3kapthe}) and the horizontal dotted line to its limit for $\theta \rightarrow 0$ (no smoothing). The slightly lower horizontal solid line shows the approximation (\ref{Spapp}). The data points are the results of numerical simulations from Jain et al.(1999). }

\label{figS3kappa}

\end{figure}

We compare in Fig.\ref{figS3kappa} our predictions to the results of numerical simulations from Jain et al.(1999) for three cosmologies. Although the numerical results probe scales which are not entirely within the non-linear regime ($\theta \ga 1'$), as shown by the variation with $\theta$ of $S_{\kapthe,3}$ which does not appear to have reached its asymptotic value at $\theta = 1'$ yet, we can check that they roughly agree with our predictions. In particular, we recover the behaviour of the dependence on the cosmology parameters $(\Om,\Ol)$. However, our predictions are somewhat larger than the numerical results obtained for low-density universes. This might be due to a dependence on $\Om$ of the parameters $S_p$ defined in (\ref{phiy}). However, as shown by Peacock \& Dodds(1996) from numerical simulations the relation between the linear and non-linear power-spectra for low-density universes is identical to the one obtained for the critical density universe after one properly takes into account the slower growth of the linear density fluctuations. That is the density and the size of matter condensations on small non-linear scales, which collapsed when $\Om \simeq 1$, is the same as the one which would be obtained in a critical density universe as seen through $\xia$. This suggests that the higher-order correlation functions and moments should also be the same which means that the coefficients $S_p$ should not depend on $\Om$ (the clustering pattern of high-density collapsed regions is expected to be roughly the same).

Moreover, we note that the coefficient $S_3$ which measures the skewness of the {\it p.d.f.} of the density contrast $\delta_R$ on small scales shows some dispersion with different numerical simulations. Indeed, while we use $S_3 \simeq 10.7$ as obtained by Valageas et al.(2000), we note that Colombi et al.(1997) find $S_3 \simeq 10.2$ and Munshi et al.(1999) get $S_3 \simeq 6$. As shown by (\ref{S3kapthe}) the skewness $S_{\kapthe,3}$ of the convergence must display the same uncertainty. Thus, it would be interesting to measure the coefficients $S_3$ and $S_{\kapthe,3}$ in the same simulation. The coefficient $S_{\kapthe,3}$ predicted by (\ref{S3kapthe}) increases for larger $\theta$ while numerical results show a decline. This decrease is mainly due to the fact that for larger angular windows the contribution from linear scales grows, which must lead to a smaller skewness $S_{\kapthe,3}$ since the coefficients $S_p$ are smaller in the quasi-linear regime (see Colombi et al.1997). The increase with $\theta$ of $S_{\kapthe,3}$ (dashed curves) is due to the oscillations of the window function $W(\Dk)$ in (\ref{IkapDelta}). Indeed, for larger angular scales $\theta$ the argument $\Dk$ for $k \sim k_{\rm m}$, where $\Delta^2(k)/k$ is maximum, reaches values of order unity so that $\Ikap$ is increasingly sensitive to the oscillations of $W(\Dk)$ (which involves the Bessel function $J_1$). As shown by Fig.\ref{figDelta2k} this appears for $\theta \ga 1'$. Then, the ratio $S_{\kaptheh,3} / S_3$ given by (\ref{S3S3mean}) grows as we get further away from the case of a constant value for the product $(w \Ikap)$ where the inequality (\ref{S3S3mean}) would transform to an equality. In this case, the ``average'' provided by (\ref{app1}) is not a very good approximation of the expression (\ref{tphikaptheh}) so that the difference between both predictions gets larger. However, we can also note that the approximation (\ref{app1}), which leads to (\ref{Spapp}), is certainly sufficient in view of the accuracy which can be obtained by such methods. The fact that the approximation (\ref{app1}) shows a better agreement with numerical results than the more precise expression (\ref{tphikaptheh}) is a mere coincidence of two effects. Indeed, our method may slightly overestimate the skewness as we do not take into account the fact that the coefficient $S_3$ gets smaller on larger scales which correspond to the mildly non-linear regime. On the other hand, as shown in Sect.\ref{Real-space method} the approximation (\ref{app1}) leads to a smaller skewness than the expression (\ref{tphikaptheh}), see (\ref{S3S3mean}). This second effect makes (\ref{app1}) to look closer to the numerical points for $\theta \ga 1'$ but on very small scales the more precise expression (\ref{tphikaptheh}) should provide the most accurate predictions.

Finally, the dependence on the angular scale $\theta$ of the skewness $S_3$ shows that for $\theta \ga 1'$ the properties of the convergence are not perfectly described by the strongly non-linear regime (also Hui 1999). Hence one would need to describe in details the transition between the highly non-linear and linear regimes, so as not to overestimate the parameters $S_p$ which are larger in the strongly non-linear regime (e.g., Colombi et al.1997). Two prescriptions have been developed to model this transition: the ``extended perturbation theory'' (Colombi et al.1997) and the ``hyper-extended perturbation theory'' (HEPT) (Scoccimarro \& Frieman 1999). Both of these empirical models use the form of the parameters $S_p$ (or of the $p-$point correlation functions) in the quasi-linear regime to propose an ``educated guess'' for their behaviour up to the strongly non-linear. They have been shown to agree reasonably well with numerical simulations. Such a description should be extended to all orders to get the {\it p.d.f.} $P(\kapthe)$ in the transitory regime, which would clearly be quite interesting. However, in this article since we focus on small angular scales we restrict ourselves to the approximation (\ref{Ikappap}) with constant coefficients $S_p$. Indeed, we shall see below in Fig.\ref{figPkappa1} and Fig.\ref{figPlkappa} that our predictions for the {\it p.d.f.} $P(\kapthe)$ agree rather well with the numerical results. This suggests that the shape of the {\it p.d.f.} $P(\kapthe)$ is not too sensitive on the value of each moment: it rather depends on the behaviour of the series of cumulants as a whole. For instance, the parameter $x_s=-1/y_s$ introduced in (\ref{ys}) and (\ref{has}) which governs the high density tail of $P(\delta_R)$ and the behaviour of $P(\kapthe)$ at large $\kapthe$, see (\ref{kapsr}), is given by:
\beq
x_s = \lim_{p \rightarrow \infty} \frac{S_{p+1}}{p S_p}
\label{xs}
\eeq
This small dependence of the {\it p.d.f.} of the density contrast $\delta_R$ (hence of the convergence $\kapthe$) on the values of the moments (except in the tails of the distribution) was already noticed in Valageas et al.(2000). Indeed, in that paper the authors found that the {\it p.d.f.} $P(\delta_R)$ obtained from various numerical simulations agree fairly well with each other (over the range tested by numerical results) while the parameters $S_p$ display a rather large dispersion, as we recalled above.

\subsection{Probability distribution}
\label{pdf kappa}

\begin{figure}

\begin{picture}(230,440)
{\epsfxsize=23 cm \epsfysize=17 cm \put(-30,-35){\epsfbox{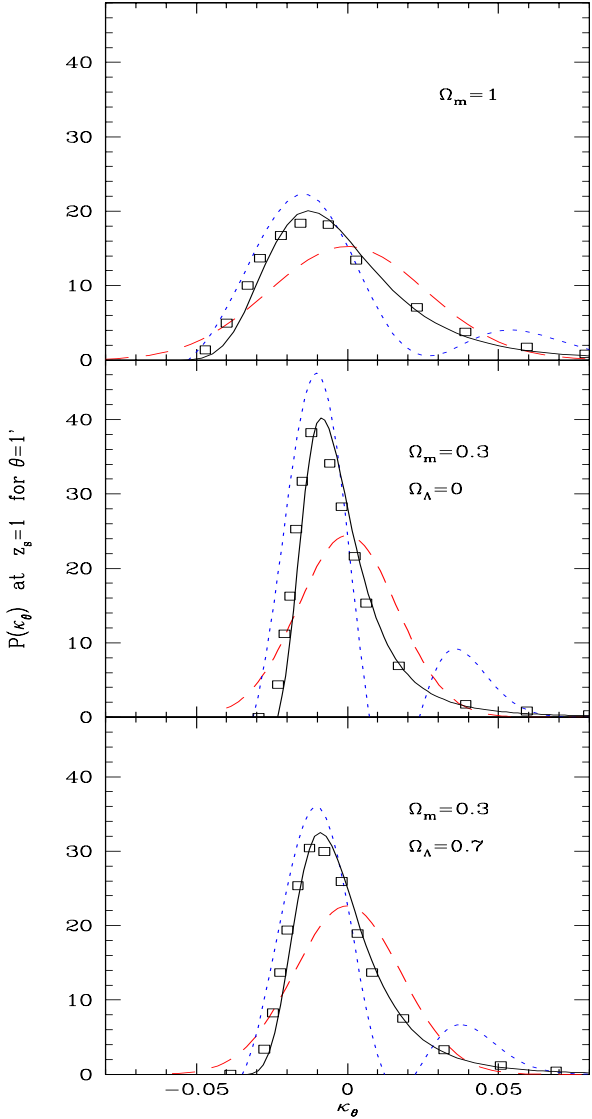}} }
\end{picture}

\caption{The {\it p.d.f.} $P(\kapthe)$ of the convergence $\kapthe$ for a source at redshift $z_s=1$ for three cosmologies and with the angular window $\theta=1'$. The solid lines show our prediction (\ref{Pkapthe}) using (\ref{Pkaptheh}) and (\ref{app1}). The dashed lines show the gaussian which has the same variance $\lag \kapthe^2 \rag$. The dotted lines correspond to the Edgeworth approximation (\ref{Edgekap}). The squares show the results of numerical simulations from Jain et al.(1999).}
\label{figPkappa1}

\end{figure}

Eventually, we can compare our predictions for the {\it p.d.f.} $P(\kapthe)$ itself with the results of numerical simulations. We show in Fig.\ref{figPkappa1} our results at $z_s=1$ for three cosmologies and the angular window $\theta=1'$. We can see that {\it our predictions} (solid lines) {\it agree very well with the numerical results} (open squares) from Jain et al.(1999). Note that at $\theta=1'$ one might have already expected a small departure due to the influence of weakly non-linear scales where the coefficients $S_p$ are slightly different from those we use which correspond to the highly non-linear regime. In particular, as seen in Fig.\ref{figS3kappa} the parameter $S_{\kapthe,3}$ we predict is not exactly equal to the one which was measured in the simulation. As discussed in the previous section, the good agreement seen in Fig.\ref{figPkappa1} means that even if our approximation for the high-order cumulants $\lag\kaptheh^p\rag_c$ is not very accurate the {\it p.d.f.} we obtain can still be a very good approximation to the actual $P(\kapthe)$. In particular, it is clear from Fig.\ref{figPkappa1} that the {\it p.d.f.} we obtain is consistent with numerical simulations for $\Om=0.3$, hence the skewness we predict, shown in Fig.\ref{figS3kappa}, must be consistent with the measure from numerical simulations. Thus, the error bars in Fig.\ref{figS3kappa} were certainly too small: they represent the dispersion of the estimator used to evaluate $S_{\kapthe,3}$ (for instance the third moment) in the simulation but other {\it p.d.f.} with a somewhat different value of $S_{\kapthe,3}$ (e.g., as large as the one we obtained) could provide a very good fit the measured $P(\kapthe)$. Note also that Fig.\ref{figPkappa1} shows that the generating function $\tvarphi(y)$ does not strongly depend on the cosmological parameters $(\Om,\Ol)$.

\begin{figure}

\begin{picture}(230,440)
{\epsfxsize=23 cm \epsfysize=17 cm \put(-30,-35){\epsfbox{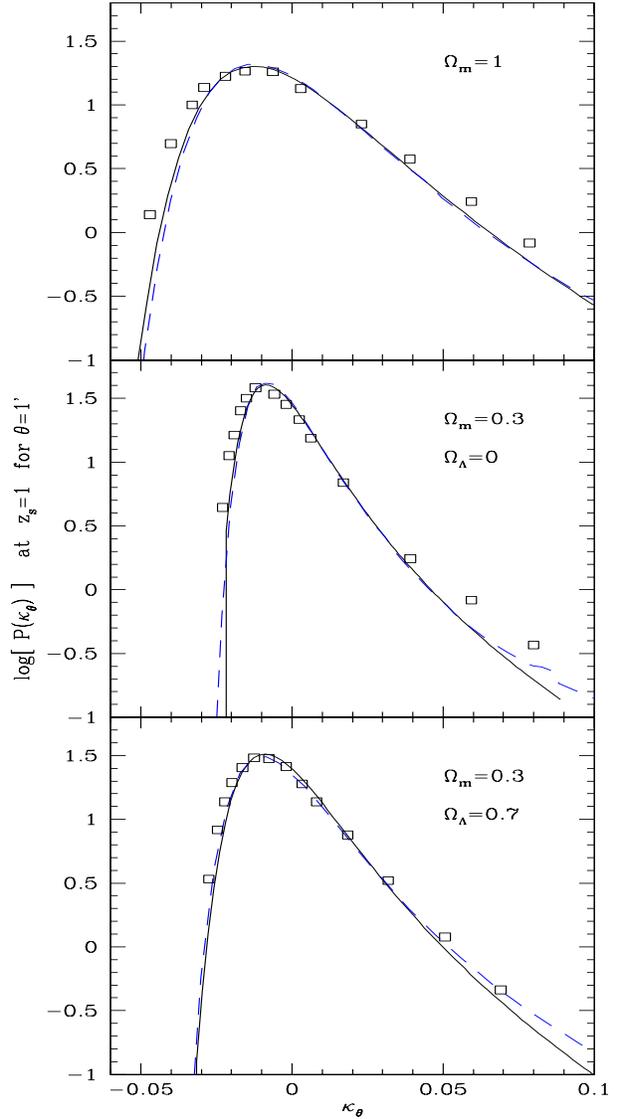}} }
\end{picture}

\caption{The logarithm of the {\it p.d.f.} $P(\kapthe)$ of the convergence $\kapthe$ for a source at redshift $z_s=1$ for three cosmologies and with the angular window $\theta=1'$. The solid lines show our predictions from the approximation (\ref{app1}) as in Fig.\ref{figPkappa1}. The dashed lines show the results obtained from (\ref{tphikaptheh}). The squares are numerical results from Jain et al.(1999).}
\label{figPlkappa}

\end{figure}

We also display in Fig.\ref{figPkappa1} the gaussian (dashed line) which would have the same variance $\lag \kapthe^2 \rag$ as our prediction. This shows that $P(\kapthe)$ cannot be approximated by a gaussian. In particular, one can clearly see the asymmetry of $P(\kapthe)$, with a maximum at a small negative value of $\kapthe$ and an extended tail at large positive $\kapthe$. Of course, this is due to the features of the {\it p.d.f.} of the density contrast itself. Indeed, on small non-linear scales most of the volume is made up of very low density regions (``voids'') which explains that $P(\kapthe)$ peaks at a negative value of $\kapthe$ (corresponding to negative density contrasts along the line of sight, see (\ref{kappa})). On the other hand, overdense collapsed matter condensations show an exponential tail at large densities which translates into the large-$\kapthe$ tail of $P(\kapthe)$.

When a {\it p.d.f.} shows a small departure from a gaussian it is customary to consider the asymptotic Edgeworth expansion (e.g., Bernardeau \& Kofman 1995). The latter is simply obtained by keeping the linear and quadratic terms in $y$ within the exponential in (\ref{Pkaptheh}) while expanding the exponential of the higher-order terms (of course for a gaussian $\tphikaptheh(y) = -y^2/2$ and there are no higher-order terms). To the first order in $\xikaptheh$ we obtain:
\begin{eqnarray}
P_E(\kaptheh) & = & \frac{e^{-\kaptheh^2/(2 \xikaptheh)}}{\sqrt{2 \pi \xikaptheh}} \; \left[ 1 + \sqrt{2 \xikaptheh} \; \frac{S_{\kaptheh,3}}{24} \; H_3\left( \frac{\kaptheh}{\sqrt{2 \xikaptheh}} \right) \right] \nonumber \\ & &
\label{Edgekap}
\end{eqnarray}
which involves the skewness (through $S_{\kaptheh,3}$) in the correction term. Here $H_3$ is the Hermite polynomial of order 3 defined by: $H_3(x)=8 x^3 - 12 x$. The {\it p.d.f.} (\ref{Edgekap}) is shown as the dotted line in Fig.\ref{figPkappa1}. We see that it captures the shift of the maximum of $P(\kapthe)$ towards negative $\kapthe$. However it fails for large $\kapthe$ where a spurious oscillation appears (because of the cubic Hermite polynomial). Note that the Edgeworth expansion is only an asymptotic expansion which is useless for $\sqrt{\xikaptheh} \ga 0.5$. Thus, in mildly non-linear regimes one cannot reconstruct the {\it p.d.f.} from the first few moments. In particular, the relation (\ref{xs}) clearly shows that even to obtain the basic features of $P(\kapthe)$ one needs the complete set of the coefficients $S_p$.

We present in Fig.\ref{figPlkappa} a comparison of the two estimates of $P(\kapthe)$ we obtained in Sect.\ref{Convergence kappa}. First, we note that both approximations give results which are very close and show a reasonable agreement with the results of numerical simulations. As discussed in Sect.\ref{Real-space method}, we can check that the approximation (\ref{app1}) leads to a slightly sharper cutoff for the exponential tail of $P(\kapthe)$ than the more precise expression (\ref{tphikaptheh}), see (\ref{ysiysm}). However, it is interesting to note that the simple approximation (\ref{app1}) provides very good results. Thus, in view of the inaccuracy introduced by the approximation (\ref{Ikappap}) and the dispersion of the actual coefficients $S_p$ (or equivalently of the function $h(x)$ defined in (\ref{hphi})) measured from counts-in-cells with different simulations the {\it p.d.f.} obtained from (\ref{app1}) is probably sufficient. To significantly improve over (\ref{app1}) one would probably need a better treatment than (\ref{Ikappap}) and a better accuracy of the measure of the {\it p.d.f.} $P(\delta_R)$ or a rigorous theoretical derivation of the coefficients $S_p$. The Fig.\ref{figPlkappa} also clearly shows the asymmetry of the {\it p.d.f.} $P(\kapthe)$, with a sharp cutoff for negative $\kapthe$, due to the lower bound $\kappamin$, and an extended exponential tail at large positive $\kapthe$. 

The fact that the simple approximation (\ref{app1}) agrees well with the numerical results is quite interesting. Indeed, it shows that from the {\it p.d.f.} $P(\kapthe)$ of the convergence $\kapthe$ we can directly derive the generating function $\tvarphi(y)$ which describes the non-linear density field at scales defined by a comoving wavenumber $k \sim 10$ Mpc$^{-1}$ (see Fig.\ref{figDelta2k}), that is for typical lengths $x \sim 0.1$ Mpc.  Then, from $\tvarphi(y)$ we can obtain the {\it p.d.f.} $P(\delta_R)$ of the density contrast using (\ref{Pdelta}). This shows that {\it the {\it p.d.f.} $P(\kapthe)$ of the convergence is a very efficient probe to obtain the {\it p.d.f.} $P(\delta_R)$ of the underlying density field}.

\section{Dependence on the redshift of the sources}
\label{Dependence on the redshift of the source}

\begin{figure}

\begin{picture}(230,440)
{\epsfxsize=23 cm \epsfysize=17 cm \put(-70,-35){\epsfbox{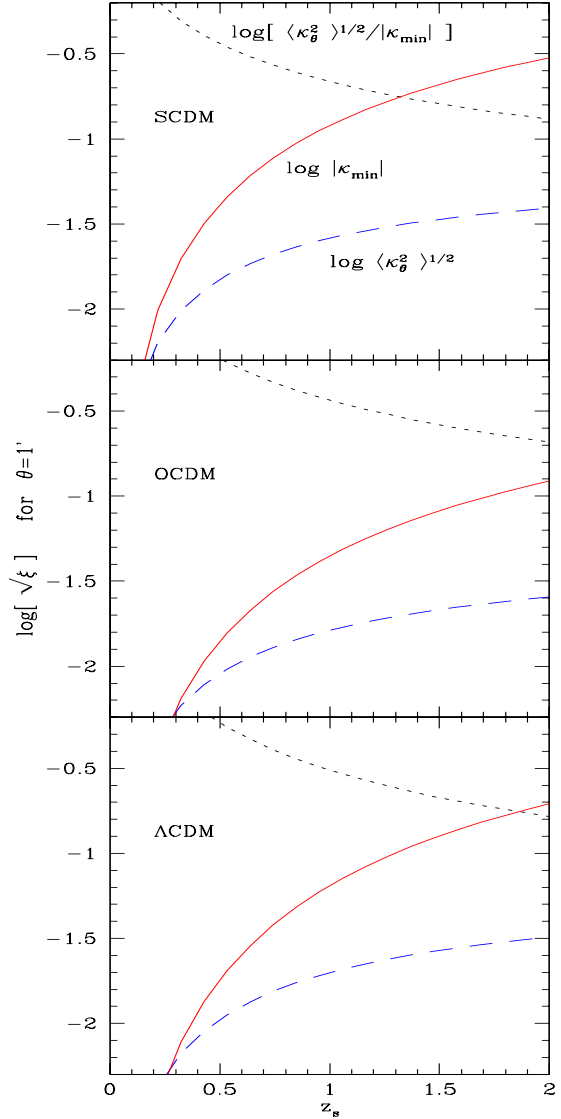}} }
\end{picture}

\caption{The dependence on the redshift $z_s$ of the source of the logarithm of the variances $\sqrt{\lag \kapthe^2 \rag}$ (convergence, dashed line) and  $\sqrt{\lag \kaptheh^2 \rag}$ (normalized convergence, dotted line), and of the lower bound $|\kappamin|$ (solid line). We consider the window $\theta=1'$ for the three cosmologies.}
\label{figXiz}

\end{figure}

\begin{figure}

\begin{picture}(230,440)
{\epsfxsize=23 cm \epsfysize=17 cm \put(-70,-35){\epsfbox{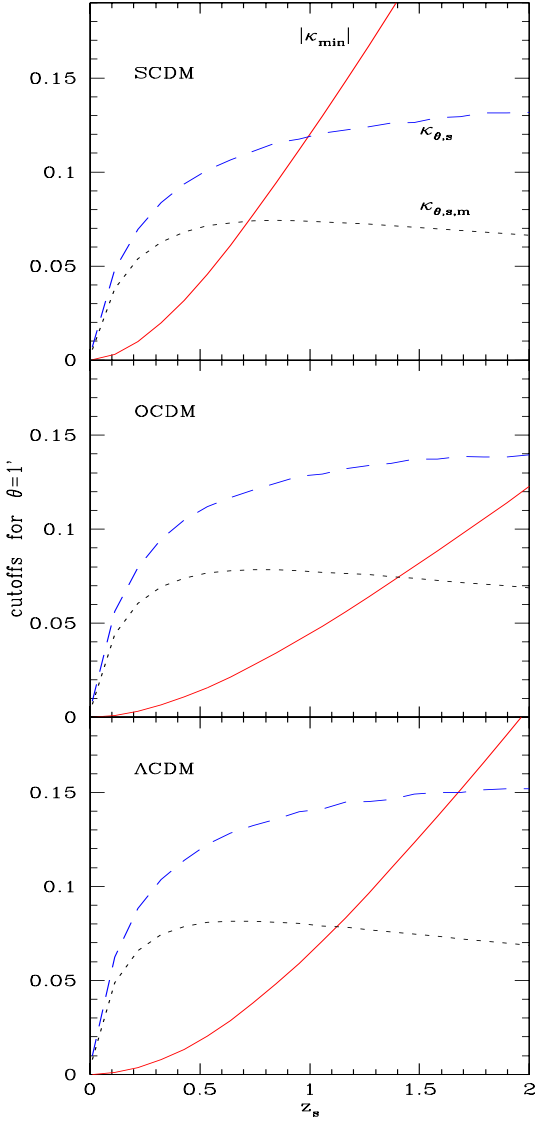}} }
\end{picture}

\caption{The dependence on the redshift $z_s$ of the cutoffs $|\kappamin|$ (solid line), $\kapso$ (dotted line) and $\kapsr$ (dashed line) for the convergence. We consider the window $\theta=1'$ for the three cosmologies.}
\label{figcutz}

\end{figure}

Finally, we briefly consider the dependence on the redshift $z_s$ of the source of the weak gravitational lensing effects we have described in the previous sections. Thus, we show in Fig.\ref{figXiz} the redshift evolution of the variance $\sqrt{\lag \kapthe^2 \rag}$ and its normalized counterpart $\sqrt{\lag \kaptheh^2 \rag}$ for the convergence $\kapthe$. First, we note that the fluctuations $\sqrt{\lag \kapthe^2 \rag}$ increase with the redshift $z_s$. Indeed, since the line of sight becomes longer as it extends to higher redshifts the distortions due to weak gravitational lensing increase. Note however that the growth of these fluctuations is rather slow beyond $z_s=1$. This is due to the slow increase of the radial coordinate $\chi$, which appears for instance in (\ref{xikapthe}), because it has a finite limit for $z \rightarrow \infty$, see (\ref{chi}). Moreover, at large redshift the density fluctuations are smaller as structure formation is less advanced, which also tends to diminish the contributions from high redshifts. On the other hand, the ``normalized'' variance $\sqrt{\lag \kaptheh^2 \rag}$ decreases for larger redshift $z_s$. Indeed, the quantity $|\kappamin|$ (solid lines) increases faster with $z_s$ than the {\it r.m.s.} fluctuation as it is not affected by the redshift dependence of the power-spectrum, see (\ref{kappamin}). In particular, as we noticed in Valageas (1999a) we obtain from Sect.\ref{Real-space method}:
\beq
z_s \rightarrow 0 \; : \; |\kappamin| \propto z_s^2 \; , \; \xikapthe \propto z_s^3 \;\;\; \mbox{and} \;\;\; \xikaptheh \propto z_s^{-1}
\eeq
and:
\beq
z_s \rightarrow \infty \; : \; \xikapthe \; \mbox{is finite and} \;\;\; \xikaptheh \propto (1+z_s)^{-2}
\eeq
Note that a smaller normalized variance $\sqrt{\lag \kaptheh^2 \rag}$ implies that the {\it p.d.f.} is closer to a gaussian near its maximum. In particular, at small redshift where $\sqrt{\lag \kapthe^2 \rag} \ga |\kappamin|$ the {\it p.d.f.} $P(\kapthe)$ is very different from a gaussian as it is strongly affected by the lower bound $\kappamin$.

Next, we display in Fig.\ref{figcutz} the evolution with redshift of the cutoff which characterizes the exponential tail $P(\kapthe)$. First, in agreement with the discussion about Fig.\ref{figXiz} we note that the magnitude $|\kappamin|$ of the lower-bound of the convergence increases faster with redshift than the exponential cutoff which depends on the normalization of the power-spectrum at each redshift, as seen for instance in (\ref{kapsr}) or (\ref{kapso}). We note that at large redshift the cutoff $\kapsr$ defined in (\ref{kapsr}) is much larger than the prediction $\kapso$, defined in (\ref{kapso}), of the simple approximation (\ref{app1}). This agrees with the discussion of Sect.\ref{Real-space method}, see (\ref{ysiysm}), and with the behaviour shown in Fig.\ref{figPlkappa} where we can see that for very large $\kapthe$ the {\it p.d.f.} $P(\kapthe)$ obtained from (\ref{app1}) is too small as compared to the prediction of the (theoretically) more accurate relation (\ref{tphikaptheh}). Nevertheless, as seen in Fig.\ref{figPlkappa} these two approximations agree very well over a large range for $\kapthe$. This shows that it is rather difficult to measure the actual cutoff of these probability distributions (see also Valageas et al.2000 for a discussion for the case of the {\it p.d.f.} $P(\delta_R)$).

As we described in Sect.\ref{Redshift distribution of the sources} our results also apply to the case of a broad redshift distribution of sources $n(z_s) dz_s$. Moreover, we noticed that since $\tvarphi(y)$ is a good approximation to $\tphikaptheh(y)$ almost all the dependence on the cosmology, the source redshift and the angular window, is enclosed in the two numbers $\kappamin$ and $\xikaptheh$ (or $\xikapthe$) (here we omit the bar used in Sect.\ref{Redshift distribution of the sources} to distinguish the case of a ``broad'' redshift distribution of sources). The variance $\xikapthe$ of the convergence can be directly obtained from the observations, hence we are left with only one parameter: $\kappamin$. Note that the latter only depends on the cosmological parameters $(\Om,\Ol,H_0)$ and on the redshift distribution of the sources, see (\ref{kappamin}): it is independent of the properties of the density field (power-spectrum,...). Thus, if these cosmological parameters are know (e.g., from CMB data and SNeIa) and the redshift of the sources is well measured it can be computed from (\ref{kappamin}). On the other hand, if one assumes that the skewness of the density field $S_3$ is known to a good accuracy (e.g., from numerical simulations) then $\kappamin$ can be obtained from the observations using (\ref{Spapp}). Bernardeau et al.(1997) advocate the use of this method to measure the cosmological parameters, using the skewness $S_3$ obtained by perturbative methods in the quasi-linear regime relevant to large angular scales ($\theta \ga 10'$), if the redshift of the sources is well measured. However, for the small angular scales studied in this article ($\theta \la 1'$) it may be better to obtain $\kappamin$ from a fit to the full {\it p.d.f.} rather than to the skewness only since the latter may not be very accurately constrained (see the discussion in Sect.\ref{Skewness}, Sect.\ref{pdf kappa} and Valageas et al.2000). Nevertheless, the observations of weak gravitational lensing effects on small scales should probably be seen as a tool to constrain the properties of the density field rather than the cosmological parameters.

\section{Conclusion}

In this article we have described a method to obtain an estimate of the {\it full {\it p.d.f.} of the convergence $\kappa$}. We summarize below the main new results we have obtained.

- Extending the method used in Valageas (1999a) we show how one can express the moments of the convergence in terms of the moments of the density field at all orders, working in real-space. This was already studied for the moments of order two and three in the non-linear regime (e.g., Hui 1999). Since {\it our approximation holds for all orders it allows us to sum up the series of the cumulants and to write the full {\it p.d.f.} $P(\kapthe)$ of the convergence in terms of the {\it p.d.f.} $P(\delta_R)$ of the density contrast} on comoving scales $x \sim 0.1$ Mpc (for angular windows $\theta \la 1'$). This provides an explicit link between two properties of the density field: the counts-in-cells statistics and the convergence of weak gravitational lensing distortions. Our results apply to the case of a broad redshift distribution of the sources as well as to the case where all sources are located at the same redshift.

- Then, we compare our predictions with the results of numerical simulations. As pointed out by previous studies (e.g., Jain \& Seljak 1997) we find that on these angular scales ($\theta \la 1'$) the non-linear evolution of the density field is important. We show that {\it our predictions for $P(\kapthe)$ agree very well with the results from N-body simulations}, although there is a non-negligible discrepancy for the skewness $S_{\kapthe,3}$. In particular, one would need a more detailed treatment (i.e. to model the transition between the highly non-linear and linear regimes) and more accurate simulations (the estimate they give for $S_3$ still varies from $6$ to $10.7$) to use the skewness on small angular scales to measure the cosmological parameters (the situation is easier at large angular scales where one can obtain $S_3$ from perturbative methods). On the other hand, it might be useful to fit the full {\it p.d.f.} rather than the skewness alone. Note that on small angular scales one cannot reconstruct $P(\kapthe)$ from its first few moments (variance and skewness) as the Edgeworth approximation is only an asymptotic expansion. Moreover the exponential tail of the {\it p.d.f.} is significantly different from a gaussian.

- We have also devised a very simple approximation which gives very good results for $P(\kapthe)$ despite its detailed properties (location of the singularity at $y_s$, value of the skewness) are somewhat different from the prediction of the more detailed prescription. It shows that {\it the measure of $P(\kapthe)$ provides a direct estimate of $P(\delta_R)$}. Moreover, it also means that the {\it p.d.f.} $P(\kapthe)$ should obey {\it a specific scaling property} to a good accuracy. Indeed, to a good approximation the {\it p.d.f.} $P(\kaptheh)$ of the normalized convergence $\kaptheh$ should only depend on one number: the variance $\xikaptheh$. Hence the {\it p.d.f.} obtained for various angular windows $\theta_1$ at a redshift $z_{s1}$ should nearly superpose onto the {\it p.d.f.} obtained for another redshift $z_{s2}$ after the rescaling such that $\xi_{\kappah_{\theta 1}} = \xi_{\kappah_{\theta 2}}$.

- Finally, we have briefly presented some aspects of the dependence of gravitational lensing distortions with the redshift of the source.

Using the relations between the properties of the density field and the gravitational lensing effects we have described in this article, one could constrain the cosmological scenario of structure formation from observations. For instance the ellipticities of images of galaxies can allow one to measure the shear $\gam$. Then, one could derive both the cosmological parameters $(\Om,\Ol)$ (e.g., through the second and third order moments) and the {\it p.d.f.} of the density field on non-linear scales. This could provide very interesting results, complementary to the usual surveys of galaxies which give indirect constraints on the mass functions. Since the latter can be related to the statistics of the counts-in-cells (e.g., Valageas \& Schaeffer 1997; Valageas et al.2000) this would provide a good check of our descriptions of the density field. We shall also present in future articles (Valageas 2000; Bernardeau \& Valageas 2000) how one can extend our results to other measures of weak gravitational lensing, like the shear or the aperture mass. 

However, we must note that in this article we did not take into account several effects which might distort the {\it p.d.f.} $P(\kapthe)$ we obtained. Thus, the non-linear coupling between deflecting lenses along the line of sight and the higher-order terms beyond the Born approximation lead to non-linear terms in the expression of the convergence as a function of the density contrast $\delta$ which would slightly change the {\it p.d.f.} $P(\kapthe)$. These effects have been studied in the quasi-linear regime (i.e. $\theta \ga 10'$) where they have been found to be negligible for low-order moments (e.g., Bernardeau et al.1997). A detailed study would be needed to estimate their magnitude in the non-linear regime but the good agreement of our predictions with the results of numerical simulations suggests that these corrections should be rather small (and probably below the inaccuracy of these simulations). On the other hand, the coupling between the sources and the lenses, which we mentionned in Sect.\ref{Redshift distribution of the sources}, is not taken into account in the numerical simulations. This was studied in the quasi-linear regime in Bernardeau (1998) who found that the correction is negligible for the skewness and the kurtosis if the width of the redshift distribution is small enough ($\Delta z_s \la 0.15$). Although we may expect similar results at small angular scales it would be interesting to investigate in details this effect.

\appendix



\section{Density probability distribution}
\label{Density probability distribution}

Here, we recall the behaviour of $P(\delta_R)$ obtained from simple forms for $h(x)$ which are consistent with numerical simulations. From very general considerations (Balian \& Schaeffer 1989) one expects the function $\varphi(y)$ defined in (\ref{phiy}) to behave as a power-law for large $y$:
\beq
y \rightarrow +\infty \; : \; \varphi(y) \sim a \; y^{1-\omega}
\hspace{0.3cm} \mbox{with} \; 0 \leq \omega \leq 1 \hspace{0.2cm} , \hspace{0.2cm} a>0
\label{phiom}
\eeq
and to display a singularity at a small negative value of $y$:
\beq
y \rightarrow y_s^+ \; : \; \varphi(y) = - a_s \; \Gamma(\omega_s)
\; (y-y_s)^{-\omega_s}
\label{ys}
\eeq
where we neglected less singular terms. From this behaviour of $\varphi(y)$ we have (Balian \& Schaeffer 1989) from (\ref{hphi}):
\beq
\left\{ \begin{array}{rl} x \ll 1 \; : & {\displaystyle  h(x) \sim
\frac{a(1-\omega)}{\Gamma(\omega)} \; x^{\omega-2} } \\ \\  x \gg 1 \; : & {\displaystyle h(x) \sim a_s \; x^{\omega_s-1} \; e^{-x/x_s} } \end{array} \right.
\label{has}
\eeq
with $x_s=1/|y_s|$. Thus, using (\ref{Phx}) we can see that the density probability distribution $P(\delta_R)$ shows a power-law behaviour from $(1+\delta_R) \sim \xia^{\;-\om/(1-\om)}$ up to $(1+\delta_R) \sim x_s \xia$ with an exponential cutoff above $x_s \xia$. At present, there is no good theoretical model to obtain the functions $h(x)$ and $\varphi(y)$. In particular, Valageas \& Schaeffer (1997) and Valageas et al.(2000) compared the functions $h(x)$ given by numerical simulations with the predictions of a simple model (consistent with the usual Press-Schechter prescription, Press \& Schechter 1974) based on the stable-clustering ansatz and the spherical collapse dynamics. They found that although this model recovers the qualitative change of $h(x)$ with the slope $n$ of the linear power-spectrum it strongly disagrees with numerical results on a quantitative level. Similar results were reached in Valageas (1998) where the full {\it p.d.f.} of the density contrast (for voids as well as for overdensities) was compared with more detailed spherical models and with the adhesion model. Thus, in this article we use the scaling function $h(x)$ obtained from numerical simulations by Valageas et al.(2000) for the case of a critical universe with an initial linear power-spectrum which is a power-law $n=-2$:
\beq
h(x) = \frac{a(1-\omega)}{\Gamma(\omega)} \; \frac{x^{\omega-2}}{(1+bx)^{c}}
\; \exp(-x/x_s)    
\label{fithx}
\eeq
with:
\[
a=1.71 \; , \; \omega=0.3 \; , \; x_s=13 \; , \; b= 5 \;\; \mbox{and} \;\; c=0.6
\]
\[ 
\mbox{hence} \;\; \omega_s = \omega -1 -c =-1.3
\]
Indeed, as seen in Fig.\ref{figDelta2k} the contributions to weak lensing distortions come from scales where $n \simeq -2$. However, the curvature of the CDM power-spectrum may slightly change the parameters $S_p$ from the value they would have for a pure power-law $P(k)$. Nevertheless, the use of (\ref{fithx}) has the advantage of the simplicity and should provide a reasonable description for all cases of interest, as seen in Fig.\ref{figPkappa1}. The scaling function $h(x)$ shown in (\ref{fithx}) defines the generating function $\varphi(y)$ through (\ref{phih}). In particular, one obtains (see Gradshteyn \& Ryzhik 1965, \S 9.211, p.1058):
\beq
\varphi'(y) = a (1-\om) b^{-\om} \; \psi \left(\om,2+\om_s;\frac{y-y_s}{b}\right)
\eeq
and
\beq
\begin{array}{l} {\displaystyle \varphi(y) = a \; b^{1-\om} \left[ \; \psi \left(\om-1,1+\om_s;\frac{y-y_s}{b}\right) \right. } \\ \\ {\displaystyle \left. \hspace{3cm} - \psi \left(\om-1,1+\om_s;\frac{-y_s}{b}\right) \right] } \end{array}
\eeq
where $\psi$ is Kummer's function which can be expressed in terms of the difference between two confluent hypergeometric functions $ _{1}F_{1}$. As emphasized in Valageas (1999a) it is better to define $\varphi(y)$ from $h(x)$ rather than trying to use a fit for $\varphi(y)$ itself. Indeed, from (\ref{phiy}) we see that:
\beq
p \geq 1 \; : \hspace{0.6cm} (-1)^{p-1} \varphi^{(p)}(0) = S_p > 0
\eeq
and moreover one can show from (\ref{Sphx}) that the coefficients $S_p$ must obey:
\beq
S_{p+q} \; S_{p-q} \geq S_p^2
\eeq
These constraints are automatically verified if one defines $\varphi(y)$ from $h(x)$. If one uses a fit for $\varphi(y)$ which does not obey these constraints one may get negative probabilities (since in this case $h(x)$ has to be negative in some range). Moreover, the expansions around $y=0$ of $\varphi(y)$ and $\tvarphi(y)$ must satisfy:
\beq
\varphi(y) = y - \frac{y^2}{2} + ... \hspace{0.3cm} , \hspace{0.3cm} \tvarphi(y) = - \frac{y^2}{2} + ...
\eeq
This is important in order to get the right location of the maximum of the various {\it p.d.f.} we consider as well as the right variance.

\section{Tree-model}
\label{Tree-model}

A popular model for the $p-$point correlation functions in the non-linear regime is to consider a tree-model (Schaeffer 1984; Bernardeau \& Schaeffer 1992) where $\xi_p$ is expressed in terms of products of $\xi_2$ as:
\beq
\xi_p({\bf r}_1, ... ,{\bf r}_p) = \sum_{(\alpha)} Q_p^{(\alpha)} \sum_{t_{\alpha}} \prod_{p-1} \xi_2({\bf r}_i , {\bf r}_j)
\label{tree}
\eeq
where $(\alpha)$ is a particular tree-topology connecting the $(p-1)$ points without making any loop, $Q_p^{(\alpha)}$ is a parameter associated with the order of the correlations and the topology involved, $t_{\alpha}$ is a particular labelling of the topology $(\alpha)$ and the product is made over the $(p-1)$ links between the $p$ points with two-body correlation functions. In this case, the coefficients $S_p$ defined in (\ref{phiy}) are given by:
\beq
S_p = \sum_{(\alpha)} Q_p^{(\alpha)} \sum_{t_{\alpha}} \frac{ \int_V \frac{d^3r_1 ... d^3r_p}{V^p} \; \prod_{p-1} \xi_2 ({\bf r}_i,{\bf r}_j) } { \left[ \int_V \frac{d^3r_1 d^3r_2}{V^2} \; \xi_2 ({\bf r}_1,{\bf r}_2) \right]^{p-1} }
\label{Sptree}
\eeq
On the other hand, the convergence $\kapthe$ involves integrals along the line-of-sight. In particular, we need the quantities $\om_p$ defined by (see (\ref{kappap2})):
\beq
\om_p({\vec \zeta}_1,...,{\vec \zeta}_p;z) = \int_{-\infty}^{\infty} \prod_{i=2}^p d\chi_i \;\; \xi_p \left( \begin{array}{l} 0 \\ \De \; {\vec \zeta}_1 \end{array} , ... ,  \begin{array}{l} \chi_p \\ \De \; {\vec \zeta}_p \end{array} ;z \right)
\eeq
Then, it is easy to see that the angular functions $\om_p$ display the same tree-structure (\ref{tree}) as the 3D correlation functions $\xi_p$ (see Bernardeau 1995 for a detailed study in the quasi-linear regime):
\beq
\om_p({\vec \zeta}_1,...,{\vec \zeta}_p;z) = \sum_{(\alpha)} Q_p^{(\alpha)} \sum_{t_{\alpha}} \prod_{p-1} \om_2({\vec \zeta}_i , {\vec \zeta}_j ;z)
\label{treeom}
\eeq
with:
\beq
\om_2({\vec \zeta}_1,{\vec \zeta}_2;z) = \pi \int_0^{\infty} \frac{dk}{k} \; \frac{\Delta^2(k)}{k} \; J_0 \left( \De k |{\vec \zeta}_1 - {\vec \zeta}_2| \right)
\label{om2}
\eeq
The cumulants $\lag\kaptheh^p\rag_c$ depend on the mean over the disk of radius $\theta$ of the quantities $\om_p$, see (\ref{Ikappap}), that is on the coefficients $s_p$ given by:
\beq
s_p = \sum_{(\alpha)} Q_p^{(\alpha)} \sum_{t_{\alpha}} \frac{ \int \frac{d^2 \zeta_1 ... d^2 \zeta_p}{(\pi \theta^2)^p} \; \prod_{p-1} \om_2 ({\vec \zeta}_i,{\vec \zeta}_j) } { \left[ \int \frac{d^2 \zeta_1 d^2 \zeta_2}{(\pi \theta^2)^2} \; \om_2 ({\vec \zeta}_1,{\vec \zeta}_2) \right]^{p-1} }
\label{sptreeom}
\eeq
Note the similarity with (\ref{Sptree}). However, since the volume averages are different (a disk or a sphere) the relative weight of the various topologies are slighlty different in (\ref{sptreeom}) as compared to (\ref{Sptree}), so that $s_p$ cannot be obtained directly from $S_p$. In (\ref{Ikappap}) we simply made the approximation $s_p \simeq S_p$ which should be quite reasonable. For instance, Gaztanaga (1994) obtains $s_3 \simeq 0.981 S_3$ for a power-law two-point correlation function $\xi(r) \propto r^{-1.8}$. Note moreover that we can expect the approximation (\ref{Ikappap}) to be reasonable for models more general than the tree-model (\ref{tree}). Furthermore, the inaccuracy introduced by the approximation $s_p \simeq S_p$ is much smaller than the uncertainty which affects the measure of $S_p$ from counts-in-cells statistics, as noticed in Sect.\ref{Skewness}.

\end{document}